\documentclass[aps,showpacs,twocolumn,superscriptaddress]{revtex4}

\newcommand{\dt}{(\partial_{t} - {\cal L}_\beta)\;}
\newcommand{\bea}{\begin{eqnarray}}
\newcommand{\eea}{\end{eqnarray}}
\newcommand{\beq}{\begin{equation}}
\newcommand{\eeq}{\end{equation}}
\newcommand{\ar}{\alpha^{'}}
\newcommand{\br}{\beta^{'}}
\newcommand{\gr}{\gamma^{'}}

\newcommand{\arr}{\alpha^{''}}

\newcommand{\epower}{\begin{rm}{e}\end{rm}}

\usepackage{amsmath}
\usepackage{amssymb}
\usepackage{latexsym}
\usepackage{graphicx}

\usepackage{graphics}
\usepackage{psfig}
\usepackage{epsfig}
\usepackage{color}
\usepackage{changebar}

\begin{document}


\title{Maximal Slicing for Puncture Evolutions of \\ Schwarzschild 
       and Reissner-Nordstr\"om Black Holes}


\author{Bernd Reimann}
\affiliation{Max Planck Institut f\"ur Gravitationsphysik,
Albert Einstein Institut, Am M\"uhlenberg 1, 14476 Golm, Germany}

\author{Bernd Br\"ugmann}
\affiliation{Center for Gravitational Physics and Geometry, Penn State
University, University Park, PA 16802, U.S.A.}


\date{July 9, 2003}


\begin{abstract}
We prove by explicit construction that there exists a maximal slicing
of the Schwarzschild spacetime such that the lapse has zero gradient
at the puncture. This boundary condition has been observed to hold in
numerical evolutions, but in the past it was not clear whether the numerically
obtained maximal slices exist analytically. We show that our
analytical result agrees with numerical simulation. Given the
analytical form for the lapse, we can derive that at late times the
value of the lapse at the event horizon approaches the value
\hbox{$\frac{3}{16}\sqrt{3} \approx 0.3248$}, justifying the numerical
estimate of $0.3$ that has been used for black hole excision
in numerical simulations. We present our results for the non-extremal
Reissner-Nordstr\"om metric, generalizing previous constructions of
maximal slices.
\end{abstract}


\pacs{
04.25.Dm,   
04.70.Bw,   
95.30.Sf    
\quad Preprint numbers: AEI-2003-058, CGPG-03/7-1
}


\maketitle


\section{Introduction}
\label{sec:introduction}

When decomposing the 4-dimensional Einstein equations into (3+1)-dimensional form, a crucial role is played by the lapse function that determines how the 4-dimensional manifold is sliced into 3-dimensional hypersurfaces.
For spatial hypersurfaces the lapse determines how time progresses between hypersurfaces. 
There is large freedom in choosing a lapse, but a particularly interesting possibility is to determine the lapse from the condition that the mean extrinsic curvature of the slices vanishes at all times. 
This condition corresponds to a certain maximal volume property of the hypersurface \cite{York79}, and the resulting gauge condition is referred to as maximal slicing.

Maximal slicing has played an important role in both analytical and numerical work using 3+1 decompositions.
It was suggested as geometrically motivated anti-focussing slicing condition by A.\ Lichnerowicz \cite{Lichnerowicz44} already in 1944, i.e.\ maximal slices avoid certain coordinate pathologies. 
Furthermore, maximal slicing helps to deal with physical singularities associated with black holes.
For a single Schwarzschild black hole, maximal slicing can be constructed analytically \cite{Estabrook73,Beig98}. 
One can obtain maximal slices that extend from the ``right-hand'' spatial infinity to the ``left-hand'' inner infinity of the extended Schwarzschild spacetime, and these slices give a complete foliation outside the event horizon while approaching a limiting slice inside that does not reach the physical singularity. In this sense maximal slices avoid the physical singularity, and even in more general situations maximal slices have been found to be singularity avoiding \cite{Eardley79}.

For these reasons, maximal slicing has been used frequently in numerical simulations of one black hole, e.g.\ \cite{Estabrook73,Bernstein89,Anninos95}, and also in binary black hole mergers in full 3D, e.g.\ \cite{Bruegmann97,Alcubierre00b}.
The lapse for maximal slicing is obtained in general by solving an elliptic equation, but even though there now are algebraic lapse conditions that are computationally far less expensive and share the singularity avoidance property, maximal slicing is still being used when smoothness of the lapse is an issue, say in the currently most advanced computations of gravitational waves from black hole mergers \cite{Baker:2001nu}. 
In addition, there are shift conditions \cite{Alcubierre02a} which in many cases overcome the so-called slice stretching problem of maximal slicing that previously limited black hole evolutions like \cite{Alcubierre00b} to short evolution times.

In this paper we establish the connection between the analytically known solutions for maximal slicing of the Schwarzschild spacetime \cite{Estabrook73,Beig98} and the maximal slicing computed in the puncture evolution method for black holes \cite{Bruegmann97}.
In the puncture method, the left-hand asymptotically flat region of a black hole is effectively compactified by analytically factoring out a coordinate singularity and working on $R^3$, which can result in a significant technical simplification over approaches which implement an inner boundary for black hole excision. 
However, although the general solution to the linear elliptic equation for the maximal slicing lapse is known for Schwarzschild, this solution depends on the boundary condition imposed on the lapse. 
Only ``odd'' and ``even'' boundary conditions have been considered before, and the numerical lapse of the puncture evolutions is not of this type. 
Furthermore, there was no rigorous proof that in the presence of the puncture coordinate singularity a regular solution to the lapse equation exists, and while numerically quite robust, it was not clear that the numerics was able to reliably determine this lapse.

The key idea in this paper is to impose the numerically observed behaviour of the lapse, namely that its first derivative vanishes at the puncture, as a novel boundary condition on the lapse. 
This boundary condition we call the ``zero gradient at the puncture'' (``zgp'') condition.
Since it is the coordinate singularity at the puncture that implies the ``zgp'' condition, we refer to the resulting maximal slicing lapse also as the puncture lapse.

The main result of the paper is that we can prove by explicit construction that there exists a maximal slicing of the Schwarzschild spacetime such that the lapse has zero gradient at the puncture, and furthermore that our analytical result agrees with numerical simulation. 
Concretely, we obtain the ``zgp'' or puncture lapse as a time-dependent linear superposition of the odd lapse and the even lapse. 
As an application of the analytical form for the puncture lapse, we can derive that at late times the value of the lapse at the right-hand event horizon approaches the value \hbox{$\frac{3}{16}\sqrt{3} \approx 0.3248$}, justifying the numerical estimate of $0.3$ that has been used for black hole excision in numerical simulations.

There are two technical difficulties to overcome. First, the analytic formulas of \cite{Estabrook73,Beig98} are non-trivial, in particular the calculation is based on Schwarzschild coordinates which lead to complications in the numerical evaluation of the resulting integrals. 
And second, we have to establish an explicit mapping between the coordinates used in our analytical study and the numerical simulation before we can perform a comparison. 
In this paper we only consider the case of vanishing shift.

Since we have to discuss \cite{Estabrook73,Beig98} in quite some detail, we take the opportunity to derive all relevant equations not only for Schwarzschild, but for a non-extremal Reissner-Nordstr\"om black hole. 
This turns out to be possible without major changes to the formalism and extends in some places previous results \cite{Duncan85}. 

The paper is organized as follows.
In Sec.~\ref{sec:generallapse}, we recall the analytic solution to the lapse equation.
In Secs.~\ref{sec:oddlapse} and \ref{sec:evenlapse}, we discuss in some detail the odd and even lapse, respectively, since these solutions will be superimposed to obtain the ``zgp'' lapse. 
Furthermore, in Sec.~\ref{sec:evenlapse} we make explicit contact between the equations given in \cite{Estabrook73} and \cite{Beig98}. 
In Sec.~\ref{sec:zgplapse}, we construct the ``zgp'' lapse and discuss the late time limit of the lapse at the right-hand event horizon. 
In Sec.~\ref{sec:compareanalytic}, we compare the analytical results for odd, even, and ``zgp'' maximal slices. 
In Sec.~\ref{sec:numericallapse}, we construct the coordinate transformation needed to make contact with the numerical ``zgp'' lapse and compare with our analytical result. 
We conclude in Sec.~\ref{sec:conclusion}.


\section{Lapse for maximal slicing of the Reissner-Nordstr\"om spacetime}
\label{sec:generallapse}
To derive a differential equation for the lapse which maximally slices the Reissner-Nordstr\"om spacetime, we follow the reference \cite{Estabrook73} closely. There in 1973 - concentrating on even boundary conditions - F.\ Estabrook and H.\ Wahlquist were the first to find and \hbox{S.\ Christensen}, \hbox{B.\ DeWitt}, \hbox{L.\ Smarr} and \hbox{E.\ Tsiang} to verify numerically the analytic solution for maximal slicing of a Schwarzschild black hole. 
However, to simplify some calculations for a comparison of different boundary conditions imposed on this lapse, as done in Sec.~\ref{sec:oddlapse} to \ref{sec:zgplapse}, the notation of \cite{Beig98} will be used to some extent and will be generalized to include electric charge. 
In the latter paper, \hbox{R.\ Beig} and N.~\'{O} Murchadha re-derived this foliation based on a more geometrical ansatz and studied for even boundary conditions the late time behaviour of the throat in the limit as proper time at infinity becomes arbitrarily large. 
We will make contact between formulas of those two references in Sec.~\ref{sec:evenlapse}.

We start with the source-free Einstein-Maxwell equations
\bea
\label{eq:EinsteinEQ}
	G_{\mu\nu} := R_{\mu\nu} - \frac{1}{2} R \: g_{\mu\nu} & = & 8\pi T_{\mu\nu} , \\
\label{eq:MaxwellEQ}
        \nabla^{\mu} F_{\mu\nu} & = & 0 ,
\eea
for a charged black hole, where the stress-energy tensor $T_{\mu\nu}$ is given
in terms of the Maxwell tensor $F_{\mu\nu}$ by
\beq    
\label{eq:stressenergy}
	T_{\mu\nu} = \frac{1}{4\pi} (F_{\mu\sigma} F_{\nu}^{\ \sigma}
				     - \frac{1}{4} g_{\mu\nu} F_{\sigma\varrho} F^{\sigma\varrho}).
\eeq
The 3+1 split of the equations in the ADM-formalism (see R.\ Arnowitt, S.\ Deser and C.W.\ Misner \cite{Arnowitt62} or \cite{York79}) yields the Hamiltonian and momentum constraints
\bea
\label{eq:Hamcon} 
	{\cal H} := & R + K^2 - K_{ij} K^{ij} - 16\pi \rho & = 0, \\
\label{eq:momcon} 
	{\cal M}_{i} := & \nabla_{j} (K_{i}^{\ j} - \gamma_{i}^{j} K) + 8\pi s_{i} & = 0 ,
\eea
together with the evolution equations
\bea
\label{eq:evm} 
	\dt \gamma_{ij} & = & - 2 \alpha K_{ij}, \\
\label{eq:evK} 
	\dt K_{ij} & = & 
 (- \nabla_{i} \nabla_{j} + R_{ij} + K K_{ij} - 2 K_{ik}{K^k}{}_{j}) \: \alpha \nonumber \\
        	 \ & \ &  - (8\pi S_{ij} - 4\pi (S - \rho)\gamma_{ij}) \: \alpha.
\eea
Here the fundamental dynamical variables are the induced, positive definite 3-metric $\gamma_{ij}$ and the extrinsic curvature $K_{ij}$.
Note that ${\cal L}_{\beta}$ is the Lie derivative with respect to the shift vector and $\nabla_i$ is the covariant spatial derivative associated with $\gamma_{ij}$. 
Furthermore, $R_{ij}$ is the 3-dimensional Ricci tensor and the Ricci scalar $R$ is its trace. 
The stress tensor $S_{ij}$, the momentum density vector $s_{i}$, and the total energy density $\rho$ are obtained as projections of the Maxwell stress-energy tensor $T_{\mu\nu}$ along the normal $n^{\mu}$ by
\beq
\label{eq:Tmunuprojection} 
	\rho = T_{\mu\nu}n^{\mu}n^{\nu}, \ 
	s_{i} = T_{i\mu}n^{\mu}, \ \ \begin{rm}{and}\end{rm} \ \ 
	S_{ij} = T_{ij}.
\eeq
The mean extrinsic curvature
\beq
\label{eq:Kofn}
	K := \gamma_{ij} K^{ij} = - \nabla_{\mu} n^{\mu}
\eeq
measures the amount of ``crunch'', i.e.\ the fractional rate of contraction of 3-volume along a unit normal to the surface. 

In the following we want to restrict ourselves to maximal slicing,
\beq
	K \equiv 0.
\eeq 
The reason for its name can be inferred from a variational
principle maximizing the volume
\beq
\label{eq:volume}
	{\cal V}(\Omega) = \int_{\Omega} \sqrt{\det{\left\{\gamma_{mn}\right\}}} \:d^{3}x
\eeq
of a bounded but arbitrary portion $\Omega$ of the Cauchy \hbox{slice $\Sigma$}, where as pointed out by J.W.\ York in \cite{York79} it follows that the trace of the extrinsic curvature has to vanish.
Contracting the evolution equation for $K_{ij}$, (\ref{eq:evK}), in the context of $K \equiv 0$ one obtains the maximality condition
\beq
\label{eq:dda=Ra} 
	\triangle \alpha := \nabla^{i} \nabla_{i} \alpha = \left( R + 4\pi (S - 3\rho) \right) \alpha,
\eeq
which is an elliptic equation for the lapse function.

As in \cite{Estabrook73,Beig98,Duncan85} the calculations are performed in the radial gauge
\beq
\label{eq:4mradial}
	ds^{2} = (-\alpha^{2}+\frac{\beta ^{2}}{\gamma})\:d\tau^{2} 
               + 2\beta\:d\tau dr 
               + \gamma\:dr^2 
               + r^{2}\:d\Omega^{2}
\eeq
with $\alpha, \beta$ and $\gamma$ being functions of $\tau$ and $r$ only. 
Whereas \hbox{$\{t,r,\theta,\phi\}$} are the standard Schwarzschild coordinates with the Schwarzschild radius $r$ measuring the circumference divided by \hbox{$2 \pi$} or the square root of area divided by \hbox{$4 \pi$} of the 2-sphere, the function \hbox{$\tau = \tau(t,r)$} is clearly not. 

Note furthermore that for the Maxwell tensor $F_{\mu\nu}$ of a purely radial electrostatic field one can make the ansatz
\beq
	F_{\mu\nu} =  \left(
				\begin{array}{cccc}
					0 & -E_{Q} & 0 & 0 \\
					E_{Q} & 0 & 0 & 0 \\
					0 & 0 & 0 & 0 \\
					0 & 0 & 0 & 0
				\end{array}
		      \right),	
\eeq
see e.g.\ Chap.~18 of \cite{Dinverno92}, and one can readily verify that 
\beq
\label{eq:E1}
	E_{Q} = \frac{Q\alpha\sqrt{\gamma}}{r^{2}}
\eeq
solves the source-free Maxwell's equation (\ref{eq:MaxwellEQ}). 
For (static) Schwarzschild coordinates the classical result \hbox{$E_{Q} = \frac{Q}{r^{2}}$} for the electric field of a point particle with charge $Q$ situated at the origin is recovered. 
Finally, the stress-energy tensor can be calculated according to (\ref{eq:stressenergy}) and its projections (\ref{eq:Tmunuprojection}) are
\beq
\label{eq:Tijproject}
	\rho  = \frac{Q^{2}}{8\pi r^{4}}, \
	s_{i} = 0, \ \ \begin{rm}{and}\end{rm} \ \ 
	S_{ij} = \rho \: \begin{rm}{diag}\end{rm}
                     \left[
			       -\gamma,
                               r^{2},
                               r^{2}\sin^{2}{\theta}
		     \right].
\eeq
In addition, we observe that the trace of $S_{ij}$ coincides with $\rho$, \hbox{$S := \gamma^{ij} S_{ij} = \rho$}.

One can readily show that the equation for the vanishing of the trace of the extrinsic curvature, \hbox{$K \equiv 0$}, can as in Eq.~(1') of \cite{Estabrook73} be written in the form
\beq
\label{eq:maxgeneral}
	\left( \log{\gamma} \right)_{,\tau} 
        = \frac{\beta}{\gamma} 
          \left( 
             \log{\left[ \frac{\beta^{2}r^{4}}{\gamma} \right]}
          \right)',
\eeq
which can be used in the following to eliminate derivatives of $\gamma$ with respect to $\tau$. 
After some rewriting, the Hamiltonian and momentum constraints, (\ref{eq:Hamcon}) and (\ref{eq:momcon}), yield
\beq
\label{eq:HKgeneral}
	r \frac{\gr}{\gamma}
	- 3\frac{\beta^{2}}{\alpha^{2}\gamma}
        + \gamma (1-\frac{1}{\gamma}-\frac{Q^{2}}{r^{2}})  = 0
\eeq
and
\beq
\label{eq:MKgeneral}
        \left( \log{\left[\frac{r^{2}\beta}{\alpha\gamma}\right]} \right)'  =  0 ,
\eeq
respectively. 
Furthermore, the ${\theta\theta}$-component of the evolution equation of the extrinsic curvature, (\ref{eq:evK}), is used together with the maximality condition (\ref{eq:dda=Ra}), which for the radial gauge reduces to a condition involving the ${rr}$- and ${\theta\theta}$-component of (\ref{eq:evK}) only, to obtain as further equations
\bea
\label{eq:EVKgeneral}
	(\log{\frac{\beta}{\alpha}})_{,\tau}
        & = & -\left(
                     \frac{\beta}{\gamma} + \frac{\alpha^{2}}{\beta}
              \right) \frac{\ar}{\alpha} 
	      +\frac{3}{\gamma} \br
	      +\frac{1}{2} \left(
                        \frac{\alpha^{2}}{\beta} - \frac{4\beta}{\gamma}
                           \right) \frac{\gr}{\gamma} \nonumber \\
      \ & \ & +\left( 3\frac{\beta}{\gamma}
                     +\frac{\alpha^{2}\gamma}{\beta}
                     -\frac{\alpha^{2}}{\beta} 
              \right) \frac{1}{r}
             -\frac{Q^{2}\alpha^{2}\gamma}{r^{3}\beta}
\eea
and
\beq
\label{eq:contrEVKgeneral}
	\frac{\arr}{\alpha} 
        + \left( \frac{2}{r} -\frac{1}{2} \frac{\gr}{\gamma} \right) \frac{\ar}{\alpha}
	- \frac{2}{r} \frac{\gr}{\gamma}
        - \frac{2\gamma}{r^{2}} (1-\frac{1}{\gamma}-\frac{Q^{2}}{2r^{2}})  = 0.
\eeq
The equations (\ref{eq:HKgeneral}) to (\ref{eq:contrEVKgeneral}) generalize the Schwarzschild black hole present in Eq.~(2') to (5') of \cite{Estabrook73} to a black hole carrying electric charge. (Note the correction of two signs in \cite{Estabrook73}.) 

The momentum constraint (\ref{eq:MKgeneral}) immediately yields 
\beq
\label{eq:betaintermsof}
	\beta(\tau,r) = \frac{\alpha(\tau,r)\gamma(\tau,r)}{r^{2}}C(\tau),	
\eeq
where $C$ is a function of $\tau$ only. This equation can be used to replace $\beta$ in the remaining equations such that from the Hamiltonian constraint a first-order partial differential equation (PDE) for $\gamma$ arises.
Its solution is given by 
\beq
\label{eq:gammafinal}
	\gamma(\tau,r)  =  \frac{1}{1 - \frac{2M}{r} + \frac{Q^{2}}{r^{2}} + \frac{C^{2}(\tau)}{r^4}} 
                        = \frac{r^{4}}{p_{C}(\tau,r)} ,
\eeq
where using the function
\beq
	f(r) = 1 - \frac{2M}{r} + \frac{Q^{2}}{r^{2}}
\eeq
for convenience the polynomial 
\bea
\label{eq:pC}
	p_{C}(\tau,r) & = & r^{4}f(r) + C^{2}(\tau) \nonumber \\ 
                    \ & = & r^{4} - 2M r^{3} + Q^{2} r^{2} + C^{2}(\tau)
\eea
has been introduced.
Note that $M$, which arises as a constant of integration, could in principle be a function of $\tau$, however as in \cite{Estabrook73} it follows from the evolution equation (\ref{eq:EVKgeneral}) together with (\ref{eq:maxgeneral}), (\ref{eq:betaintermsof}) and (\ref{eq:gammafinal}) that $M$ must be $\tau$-independent.
It is worth pointing out that the radial metric function we obtain in (\ref{eq:gammafinal}) generalizes Eq.~(9) of \cite{Estabrook73} by the charge, and the same 3-metric for maximal slicing of Reissner-Nordstr\"om as found by \hbox{M.\ Duncan} in \cite{Duncan85} is obtained. 
 
Finally, making use of $\beta$, (\ref{eq:betaintermsof}), and $\gamma$, (\ref{eq:gammafinal}), the maximal slicing condition (\ref{eq:maxgeneral}) reduces to
\beq
\label{eq:differentiallapse}
	(\alpha\sqrt{\gamma})' =  - \frac{\gamma^{3/2}}{r^{2}} \frac{dC}{d\tau},
\eeq
which is the differential equation for the lapse we were looking for. 
By integration its general solution is found to be
\beq
\label{eq:generalalpha}
	\alpha(\tau,r) = \frac{\sqrt{p_{C}(\tau,r)}}{r^{2}}
                    \left( D(\tau) + \frac{dC}{d\tau} 
                                     \int^{\infty}_{r} 
                                     \frac{y^{4}\:dy}{p_{C}(\tau,y)^{\frac{3}{2}}}
                    \right)
\eeq
where $C$ and $D$ are functions of $\tau$ only and one integration limit has been fixed without loss of generality at spatial infinity. Three different boundary conditions leading to an odd, even, and puncture lapse will be discussed next.


\section{Odd Lapse}
\label{sec:oddlapse}

\subsection{Derivation}
The simplest special case of the lapse (\ref{eq:generalalpha}) is, because of its underlying antisymmetry with respect to the throat, referred to as odd (in some references also as antisymmetric) and corresponds to the boundary conditions
\beq
\label{eq:BCodd}
	\lim_{r \to \infty} \alpha^{\pm}_{odd} = \pm 1 \ \ \forall \tau_{odd}.
\eeq
To measure proper time at infinity (on the right-hand side of the throat denoted by a superscript ``$+$'') the lapse is unity there and implied by the antisymmetry the lapse is minus one at the puncture (on the left-hand side and hence denoted by a ``$-$'').
The odd lapse 
\beq
\label{eq:alphaoddfinal}
	\alpha^{\pm}_{odd}(C,r) = \pm \sqrt{f(r) + \frac{C^{2}}{r^{4}}} 
                                = \pm \frac{\sqrt{p_{C}(r)}}{r^{2}}
\eeq
is obtained by setting \hbox{$D(\tau_{odd}) = \pm 1$} and \hbox{$\frac{dC}{d\tau_{odd}} = 0$}. The time-independent slice label $C$ can be chosen independent of the time at infinity $\tau_{odd}$ and is, as pointed out for Schwarzschild in Appendix A of \cite{Beig98}, purely gauge for Reissner-Nordstr\"om.

\subsection{Schwarzschild lapse} 
The outer event horizon and inner Cauchy horizon at \hbox{$r_{\pm} = M \pm \sqrt{M^{2}-Q^{2}}$} are given by the real roots of $f(r)$, which are time independent, while the throat $r_{C}$ is obtained as a real root of $p_{C}(r)$ and therefore depends on the choice of $C$.
Hence, starting with the throat in the same location as the event horizon implies \hbox{$C = 0$}. In this case one obtains from (\ref{eq:alphaoddfinal}) the so-called Schwarzschild lapse
\beq
\label{alphastandard}
	\alpha^{\pm}_{odd} (C = 0,r) = \pm \sqrt{f(r)},
\eeq
from (\ref{eq:gammafinal}) the radial component of the metric
\beq
	\gamma(r) = \frac{1}{f(r)} ,
\eeq
and from (\ref{eq:betaintermsof}) a vanishing shift. 
Hence the static Reissner-Nordstr\"om metric written in Schwarzschild coordinates,
\beq
\label{eq:RNradial}
	ds^{2} = -f(r)\:dt^{2} + \frac{1}{f(r)}\:dr^{2} + r^{2}d\Omega^{2},
\eeq
is recovered identifying $Q$ with the charge and $M$ with the mass of the black hole \hbox{($0 \leq \left| Q \right| < M$)}. 

This result was of course expected, since it is well-known that with this coordinate choice the spacelike slices given by \hbox{$t=\begin{rm}{const}\end{rm}$}, being perpendicular to the timelike Killing vector in the exterior regions, are maximal. That one can immediately see from the definition of the extrinsic curvature $K_{ij}$, (\ref{eq:evm}), which for zero shift reduces to
\beq
\label{eq:Kijsimple}
	K_{ij} = - \frac{1}{2\alpha} \dot{\gamma_{ij}} \ \ \begin{rm}{if}\end{rm} \ \ \beta \equiv 0.
\eeq
Since the 3-metric $\gamma_{ij}$ is time-independent, each of the components and therefore also the trace of the extrinsic curvature vanishes, hence \hbox{$K \equiv 0$}. 
Note furthermore that by contraction of (\ref{eq:Kijsimple}) setting \hbox{$K \equiv 0$} one immediately obtains the statement that for maximal slicing with vanishing shift the determinant of the 3-metric is time-independent. 
Hence the singularity avoiding property comes to light as the variation of the local volume remains fixed \cite{Choptuik86}. 

\subsection{Isotropic lapse}
By applying for vanishing shift a purely spatial coordinate transformation \hbox{$r = r(R)$}, the hypersurfaces obviously remain maximal. 
To obtain isotropic coordinates by writing \hbox{$dr^{2} = (\frac{dr}{dR})^{2} dR^{2}$} and by comparing the $dt^{2}$-, $dr^{2}$- and $d\Omega^{2}$-terms the set of equations
\bea
\label{eq:TF1}
	\alpha^{\pm\:2}_{odd} (C=0,R) & = & f(r(R)) , \\ 
\label{eq:TF2}
	\Psi^{4}(R) & = & \frac{1}{f(r(R))} (\frac{dr}{dR})^2 , \\
\label{eq:TF3}
	 R^{2} \Psi^{4}(R) & = & r(R)^{2} ,
\eea
is read off. From (\ref{eq:TF2}) and (\ref{eq:TF3}) the ordinary differential equation (ODE)
\beq
\label{eq:drdR}
	\frac{dr}{dR} = \pm \sqrt{f(r(R))} \:\frac{r(R)}{R}
\eeq
can be inferred where $r$ can take values \hbox{$r\geq r_{+}$} only.
By integration the solution is found to be given by 
\bea
\label{eq:rofR}
     r(R) & = & R \left(
                  (1+\frac{M}{2R})^{2} 
                  - \frac{Q^{2}}{4R^{2}} 
                  \right) \nonumber \\
	  & = & R + M + \frac{M^{2} - Q^{2}}{4R},
\eea
choosing an arbitrary constant of integration such that $r$ and $R$ coincide at infinity. 
With \hbox{$R_{+} = \frac{1}{2}\sqrt{M^{2}-Q^{2}}$} corresponding to $r_{+}$ this relation can also be piecewise inverted to yield
\beq
\label{eq:Rofr}
\begin{array}{cc}
    R(r) = \left\{ \begin{array}{c}
			\frac{1}{2}(-M + r(1- \sqrt{f(r)})) \\
			\frac{1}{2}(-M + r(1+ \sqrt{f(r)})) 
		    \end{array}
	    \right. & \ 
        \begin{rm}{for}\end{rm} \ \ 	
	    \left\{ \begin{array}{c}
			0 \leq R \leq R_{+} \\
                        R \geq R_{+}. 
		    \end{array}
	    \right.
\end{array}
\eeq
From equation (\ref{eq:TF1}) follows the isotropic lapse 
\beq
\label{eq:alphaisotropic}
       \alpha^{\pm}_{odd}(C=0,R) = \frac{(2R+M)(2R-M) + Q^{2}}{(2R+M)^2 - Q^{2}}
\eeq
and from (\ref{eq:TF3}) one can read off the conformal factor 
\beq
\label{eq:conformalgamma}
   \Psi^{4}(R) = \left(
		  (1+\frac{M}{2R})^{2} 
                  - \frac{Q^{2}}{4R^{2}} 
                 \right)^{2}.
\eeq
Hence the Reissner-Nordstr\"om metric in isotropic coordinates as found e.g.\ in \cite{Graves60} is recovered which for vanishing charge reduces to the well-known Schwarzschild case.

\subsection{Isometry}
For the case \hbox{$0 \leq \left| Q \right| < M$}, one can see from the coordinate transformation (\ref{eq:rofR}) that there is an isometry present, since the value of $r$ and therefore also the line element (\ref{eq:RNradial}) remains the same under a mapping
\beq 
\label{eq:isometry}
	R \longleftrightarrow \frac{M^{2} - Q^{2}}{4R}. 
\eeq
These two values of $R$ for every value of \hbox{$r>r_{+}$} result in two isometric parts of the spacetime (in terms of Kruskal-Szekeres diagrams the regions I and I' in Fig.~\ref{fig:SSKruskal} for Schwarzschild). 
In particular, \hbox{$R = 0$} is simply a compactified image of infinity in the other universe referred to as puncture.
The fixed point set at the event horizon $r_{+}$ is a minimal 2-sphere of an Einstein-Rosen bridge \cite{Einstein35} located at $R_{+}$, the so-called throat (see the embedding diagram of Fig.~\ref{fig:SSembedding}).

Note also that the odd lapse in isotropic coordinates, (\ref{eq:alphaisotropic}), changes its sign with positive values on the right-hand side of the throat being the original unextended space and negative ones on the left-hand additional region obtained by analytic extension of the spacetime. In particular, the odd lapse vanishes on the throat, i.e.\ \hbox{$\alpha^{\pm}(C=0,R_{+}) = 0$} (c.f.\ the lapse profile as shown for Schwarzschild in the upper plot of Fig.~\ref{fig:SSlapseloop}).


\section{Even lapse}
\label{sec:evenlapse}

\subsection{Derivation I (via Einstein-Maxwell equations)}
The even lapse is symmetric with respect to the throat and its boundary conditions are usually stated as unit lapse at infinity and a vanishing gradient at the throat. 
Since by symmetry time has to run equally fast at both spatial ends of the manifold, one can also formulate even boundary conditions as
\beq
\label{eq:BCeven}
	\lim_{r \to \infty} \alpha^{\pm}_{even} = 1 \ \ \forall \tau_{even}.
\eeq
This motivates for the differential equation obtained for the lapse in (\ref{eq:generalalpha}) the particular choice \hbox{$D(\tau_{even}) = 1$} made by F.\ Estabrook et al.\ and generalized here, so 
\beq
\label{eq:alphaevenEsta}
	\alpha^{\pm}_{even}(C,r) = \frac{\sqrt{p_{C}(r)}}{r^{2}}
                    	      \left( 1 + \frac{dC}{d\tau_{even}} 
                                   \int^{\infty}_{r} 
                                   \frac{y^{4}\:dy}{p_{C}(y)^{\frac{3}{2}}}
                              \right). 
\eeq
Since one starts with the 3-geometry of the \hbox{$t = \tau_{even} = 0$} hypersurface, i.e.\ with the initial radial component of the 3-metric given by \hbox{$\gamma = \frac{1}{f(r)}$}, one can infer from (\ref{eq:gammafinal}) that \hbox{$C(\tau_{even} = 0) = 0$} initially. 
The function $C(\tau_{even})$ though is still undetermined as only single sheets of the 3-geometry labeled by $C$ have been looked at so far.
As stated in \cite{Estabrook73}, the $\tau_{even}$-dependence of $C$ can be fixed by imposing the requirement of smoothness across the Einstein-Rosen bridge by passing to the Reissner-Nordstr\"om line element (\ref{eq:RNradial}) in Schwarzschild coordinates.
For the ``height function'' \hbox{$t = t(C(\tau_{even}),r)$}, comparing $d\tau_{even}^{2}$-, $d\tau_{even} dr$- and $dr^{2}$-terms of the metric the following two PDEs, 
\bea
\label{eq:Estadtdtau}
	\frac{\partial t}{\partial \tau_{even}} 
		& = & \alpha_{even}^{\pm} \frac{r^{2}}{\sqrt{p_{C}(r)}} , \\
\label{eq:Estadtdr}
	\frac{\partial t}{\partial r} 
		& = & - \frac{C}{f(r)\sqrt{p_{C}(r)}} ,
\eea
are found.
Making for (\ref{eq:Estadtdr}) the ansatz 
\beq
\label{eq:EstatofCandr}
	t := t_{even}(C,r) = - \int^{r}_{s_{C}} \frac{C\:dy}{f(y)\sqrt{p_{C}(y)}},
\eeq
one finds that this equation also satisfies (\ref{eq:alphaevenEsta}) and (\ref{eq:Estadtdtau}) for $s_{C}$ coinciding with $r_{C}$. 
The latter is the unique (double counting for \hbox{$C = C_{lim}$}) larger real root of the polynomial $p_{C}(r)$ in \hbox{$\left[ r_{C_{lim}}, r_{+} \right]$} and determines the radial coordinate of the throat on a given maximal slice. 
Hence
\beq
\label{eq:tevenfinal}
	t_{even}(C,r) = H_{C}(r),
\eeq
where we have introduced the integral 
\beq
\label{eq:Hintegral}
	H_{C}(r)  := - \int^{r}_{r_{C}} \frac{C\:dy}{f(y)\sqrt{p_{C}(y)}}
\eeq
for \hbox{$r \geq r_{C}$}.
Note that in (\ref{eq:Hintegral}) the integration across the pole at $r_{+}$ is taken in the sense of the principal value, and by the same arguments as in \cite{Beig98} the corresponding slices extend smoothly through the event horizon $r_{+}$ and throat $r_{C}$. 

The procedure of deriving (\ref{eq:tevenfinal}) with (\ref{eq:Hintegral}) is non-trivial as pointed out in detail in \cite{Estabrook73}: 
The requirement of smoothness across the bridge is \hbox{$t_{even} \to 0$} and \hbox{$\frac{\partial}{\partial r} t_{even} \to \infty$} as $r$ approaches its smallest value $r_{C}$ at the center of the bridge, or in terms of Kruskal-Szekeres coordinates \hbox{$\frac{\partial}{\partial X} T =0$} as \hbox{$X \to 0$}. 
As mentioned in \cite{Geyer95}, this fixes the value of $C(r_{C})$ such that \hbox{$p_{C}(r_{C}) = 0$} holds.
In particular, with $r_{C}$ becoming at infinite times a double counting root $r_{C_{lim}}$ with the value
\beq
	r_{C_{lim}} = \frac{1}{4} (3M + \sqrt{9M^2 - 8Q^2}) > 0
\eeq
corresponding to
\bea
	C_{lim} & := & C(\tau_{even}=\infty) \\ \nonumber 
		 & = & \frac{\sqrt{2}}{8} 
		       \sqrt{27-36\frac{Q^{2}}{M^{2}}+8\frac{Q^{4}}{M^{4}}+(9-8\frac{Q^{2}}{M^{2}})^{3/2}}M^{2},
\eea
the singularity avoidance of the maximal slicing of the extended Reissner-Nordstr\"om spacetime as proven in corollary 3.8 in \cite{Eardley79} becomes clear.

Finally, since $t_{even}(r)$ and $\tau_{even}$ have to coincide at spatial infinity modulo an arbitrary constant, which can be set to zero, from (\ref{eq:tevenfinal}) in the limit \hbox{$r \to \infty$} one can infer
\beq
\label{eq:tauevenfinal}
	\tau_{even} (C) = \lim_{r \to \infty} t_{even}(C,r) = H_{C}(\infty).
\eeq

We want to stress that in (\ref{eq:EstatofCandr}) the particular choice of the larger root of $p_{C}(r)$ has been made since we are interested in the ``horizon-horizon'' subfamily of maximal slices only, which extend to both spatial infinities.
Here the slices start with the initial hypersurface representing the 3-geometry at the moment of time symmetry when the throat is at the event horizon $r_{+}$ and develop inward to approach $r_{C_{lim}}$ asymptotically and hence avoid the singularity.
As pointed out in remark 8 of \cite{Estabrook73} and in Sec.~III of \cite{Duncan85}, a second ``singularity-singularity'' subfamily of maximal slices can be obtained choosing the unique (double counting for \hbox{$C = C_{lim}$}) minor real root $\hat{r}_{C}$ of $p_{C}(r)$ in \hbox{$\left[ r_{-},r_{C_{lim}} \right]$} instead. 
The initial hypersurface is now the time-symmetric slice with the throat at the Cauchy horizon $r_{-}$ and the slices intersecting the singularity at \hbox{$r=0$} grow outward to the limiting surface $r_{C_{lim}}$. 
As discussed in the next subsection, one can compare the situation with the mechanical analogue of a particle of energy \hbox{$E = C^{2}$} travelling in a repulsive potential \hbox{$V(r) = - r^{4}f(r)$}, where $r_{C}$ and $\hat{r}_{C}$ represent turning points (c.f.\ Fig.~\ref{fig:energy}).
Thinking of particles with energies ranging from \hbox{$E = 0$} to \hbox{$E_{lim} = V(r_{C_{lim}}) = C_{lim}^{2}$} it is clear that both subfamilies together cover the spacetime with spacelike maximal slices (c.f.\ the Kruskal-Szekeres diagram for Schwarzschild and the Carter-Penrose diagram for Reissner-Nordstr\"om in Fig.~1 of \cite{Estabrook73} and \cite{Duncan85}, respectively). 
For its limited use in numerical relativity, however - as the slices lie always completely inside the event horizon and hit the singularity - we will not investigate the ``singularity-singularity'' subfamily further.

\subsection{Contact between \cite{Estabrook73} and \cite{Beig98} by a mechanical analogue}
\label{subsec:mechanics}
To make contact between formulas found by \hbox{F.\ Estabrook} et al.\ \cite{Estabrook73} and those derived by \hbox{R.\ Beig} and \hbox{N.~\'{O} Murchadha} \cite{Beig98}, one can apply an energy conservation equation worked out in Appendix B of the latter paper. 
There two functions $F$ and $J$ are introduced by
\bea
	F(E,r) & = & \int^{r}_{x_{E}} \frac{W(y)\:dy}{\sqrt{E-V(y)}}, \\
        J(E,r) & = & \int^{r}_{x_{E}} \sqrt{E-V(y)}V(y)W(y) \:dy.  \ \ \ \ \ 
\eea

In the following, $W$ may have a simple pole at \hbox{$x=\bar{x}$} and $V$ mapping $x_{E}$ to $E$ and $\bar{x}$ to $0$ is a smooth function \hbox{$V: \left[x_{0},\infty\right[ \to R$} satisfying \hbox{$0 < E < V(x_{0})$}, \hbox{$V(x)<E$}, and \hbox{$V'(x) < 0$} for \hbox{$x > x_{0}$}. 
By differentiating $J$ twice with respect to $E$ the formula
\bea
\label{eq:BeigAppendixMagic}
      4 \frac{\partial^{2}}{\partial E^{2}} J(E,x) & = & - \frac{2}{\sqrt{E-V(x)}} 
                                                           \frac{V(x)W(x)}{V'(x)} \nonumber \\
                                                 \ & \ & + \int_{x_{E}}^{x} 
                                                           \frac{2}{\sqrt{E-V(y)}} \frac{d}{dy} 
                                                           \left[ 
								\frac{V(y)W(y)}{V'(y)}
                                                           \right] dy \nonumber \\
                                                 \ & = &   F(E,x)
                                                         + 2 E \frac{\partial}{\partial E} F(E,x)
\eea
is derived, c.f.\ Eqs.~(B7) and (B9) in \cite{Beig98}. 

\begin{figure}[!ht]
	\noindent
	\epsfxsize=90mm \epsfysize=55mm \epsfbox{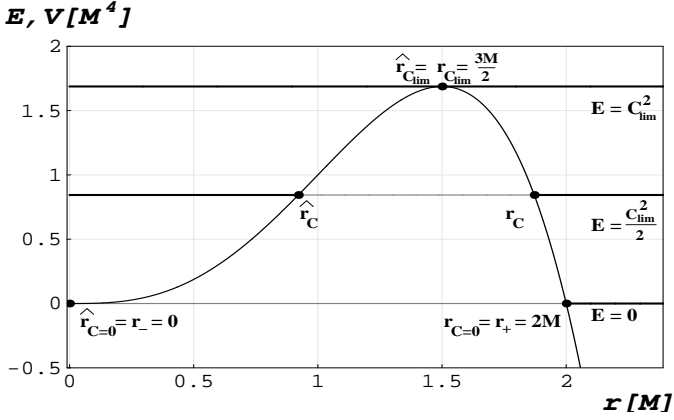} 
	\caption{For maximal slicing of Schwarzschild the mechanical analogue consisting of a particle of energy \hbox{$E = C^{2}$} travelling in a repulsive potential \hbox{$V = -  r^{4} f(r)$} is shown. For the energy levels \hbox{$E = \Big\{ 0,\frac{^{0}C_{lim}^{2}}{2},$$^{0}C_{lim}^{2}\Big\}$} the innermost and outermost radial distances for unbounded and bounded particles denoted by $^{0}r_{C}$ and $^{0}\hat{r}_{C}$, respectively, can be inferred.}
	\label{fig:energy}
\end{figure}

Now, as pointed out in \cite{Gentle2001} using proper velocity $\frac{ds}{d\tau}$, i.e.\ the rate of change of proper time along a maximal hypersurface with proper distance, from
\beq
        E = C^{2} = C^{2} \left( \frac{ds}{d\tau} \right)^{2} - r^{4} f(r) = T + V
\eeq
an energy conservation equation arises for a particle of total energy \hbox{$E = C^{2}$} moving in the repulsive potential \hbox{$V = -  r^{4} f(r)$} with a kinetic energy \hbox{$T = C^{2} \left( \frac{ds}{d\tau} \right)^{2}$}. 
By demanding \hbox{$\frac{ds}{d\tau} = 0$} for unbounded particles starting from infinity the closest approach to the singularity - or for bounded ones starting from the singularity the outermost radial distance - is found to be given by the two real roots $r_{C}$ and $\hat{r}_{C}$ of the polynomial $p_{C}(r)$, c.f.\ Fig.~\ref{fig:energy}. 
Furthermore, R.\ Beig and N.~\'{O} Murchadha \cite{Beig98} point out that for \hbox{$E > E_{lim}$} with \hbox{$E_{lim} = V(r_{C_{lim}}) = C_{lim}^{2}$} maximal slices hitting the singularity at $r = 0$ are obtained \footnote{We will not discuss these slices further. However, note that similar ``singularity-horizon'' slices extending from the singularity to infinity can be derived for constant mean curvature slicing, i.e.\ \hbox{$K \equiv$ const $\neq 0$}, see \cite{Gentle2001} for details.}. 
They also argue that $F(E,r)$ can be thought of as the time it takes a particle of energy $E$ to travel from $r_{C}$ to $r$. 
In particular, in an analysis following in a later paper \cite{mypaper2} based on \cite{Beig98, mythesis} we will be interested in the way this function blows up as $E$ approaches with $C_{lim}^{2}$ the maximum of the potential in order to study the late time behaviour of the slices. 
\pagebreak

One can readily check that the identifications
\beq
\begin{array}{ccc}
	W(r) = - \frac{1}{f(r)}, & V(r) = - r^{4} f(r), & E = C^{2}
\end{array}
\eeq
and
\beq
\begin{array}{cccc}
	r_{0} = r_{C_{lim}}, & r_{E} = r_{C}, & \bar{r} = r_{+}                 
\end{array}
\eeq
satisfy the conditions on $W$, $V$, and $E$ as stated above also in the Reissner-Nordstr\"om case and lead to
\beq
	H_{C}(r) = C F(C^2,r). 
\eeq
Hence for the derivative of $H_{C}(r)$ with respect to $C$ the important formula
\bea
\label{eq:dCdH}
	\frac{\partial}{\partial C} H_{C}(r) & = & F + 2C^{2} \frac{\partial}{\partial C^{2}} F \nonumber \\
					     & = & \frac{r^{2}}
                                                        {2(r - \frac{3}{2}M + \frac{Q^{2}}{2r}) \sqrt{p_{C}(r)}}
                                                 - \frac{1}{2} K_{C}(r) \ \ \ \ \ \ \ \ \
\eea
can be derived by straightforward calculation using (\ref{eq:BeigAppendixMagic}), where for \hbox{$r \geq r_{C}$} we have introduced the integral
\beq
\label{eq:Kintegral}
 	K_{C}(r) := \int^{r}_{r_{C}} \frac{y(y-3M+\frac{3Q^{2}}{2y})\:dy}
		                          {(y - \frac{3}{2}M + \frac{Q^{2}}{2y})^{2}\sqrt{p_{C}(y)}}. \ \ \ 
\eeq
Note that $K_{C}(r)$ diverges in the limit \hbox{$C \to C_{lim}$} as $r_{C_{lim}}$ becomes a double root of $p_{C}(r)$, the square root of which appears in the denominator of the integrand.
Note furthermore that from (\ref{eq:dCdH}) in the limit \hbox{$r \to \infty$} the equation
\beq
\label{eq:HisKhalf}
	\frac{d}{dC} H_{C}(\infty) = - \frac{1}{2} K_{C}(\infty)
\eeq
is found, c.f.\ Eq.~(B12) in \cite{Beig98}.

In order to write the expression (\ref{eq:alphaevenEsta}) for the even lapse in terms of the integral $K_{C}(r)$, one can readily verify that
\bea
\label{eq:dCdHmHisI}
	\frac{\partial}{\partial C}\left( t_{even}(C,r) - \tau_{even}(C)\right) 
     & = & \frac{\partial}{\partial C}\left( H_{C}(r) - H_{C}(\infty) \right) \nonumber \\
     & = & \int^{\infty}_{r} 
           \frac{y^{4}\:dy}{p_{C}(y)^{\frac{3}{2}}}  
\eea
holds and hence the even lapse can be written as
\bea
\label{eq:alphaevenfinal}
          \alpha^{\pm}_{even}(C,r) & = & \frac{\sqrt{p_{C}(r)}}{r^{2}}  
			 \frac{dC}{d\tau_{even}} 
			 \frac{\partial t_{even}}{\partial C} \\
                                 \ & = & 
                         - \frac{1}{K_{C}(\infty)}
                           \left( \frac{1}{r - \frac{3}{2}M + \frac{Q^{2}}{2r}} \right. \nonumber \\
                                 \ & \ & \ \ \ 
                           \left. - \frac{\sqrt{p_{C}(r)}}{r^{2}} K_{C}(r) \right), \nonumber
\eea
which generalizes Eq.~(2.23) and (2.24) in \cite{Beig98} by the charge. 
By the same arguments as applied for Schwarzschild in this reference, the term \hbox{$\frac{dC}{d\tau_{even}}\frac{\partial t_{even}}{\partial C}$} in the first line of equation (\ref{eq:alphaevenfinal}) blows up at the throat, i.e.\ for \hbox{$r \to r_{C}$}, in such a way that the lapse has a smooth limit there. 
To our knowledge the lapse function formulated in (\ref{eq:alphaevenEsta}) or (\ref{eq:alphaevenfinal}) for maximal slicing Reissner-Nordstr\"om (i.e.\ with electric charge included) has not appeared in the literature yet.

For Schwarzschild, the even lapse profile in terms of the Schwarzschild radial coordinate $r$ for the times at infinity \hbox{$\tau_{even} = \{0,1M,2M,4M,8M\}$} is shown in Fig.~\ref{fig:SSlapseloop}.

Recalling that the shift is given by (\ref{eq:betaintermsof}) and the radial component of the metric by (\ref{eq:gammafinal}), with
\beq
\label{eq:betaevenfinal}
        \beta_{even} (C,r) = \frac{\alpha_{even}(C,r)\gamma_{even}(C,r)}{r^{2}} C
\eeq
and 
\beq 
\label{eq:gammaevenfinal}
	\gamma_{even} (C,r) = \frac{1}{f(r) + \frac{C^{2}}{r^{4}}}
                            = \frac{r^{4}}{p_{C}(r)}
\eeq
the full 4-metric is found.

\subsection{Derivation II (by a variational principle and/or via construction from Reissner-Nordstr\"om metric)}

In this Subsection we present two alternative derivations of the maximal slices, which the reader may skip without loss of continuity.

Note that the maximal slicing condition $K \equiv 0$ fixes the PDE for \hbox{$\frac{\partial t}{\partial r}$} as in (\ref{eq:Estadtdr}) only, whereas boundary conditions have to be specified to obtain  $t(C,r)$ by integration. The latter can always be written as the sum of the integral $H_{C}(r)$, (\ref{eq:Hintegral}), and a time translation function of $C$, fixed initially by demanding that for \hbox{$C = 0$} one starts with the time-symmetric \hbox{$t = 0$} hypersurface, and determined during the evolution by boundary conditions. 
In particular, for even boundary conditions the throat $r_{C}$ has to stay at \hbox{$t = 0$} because of the symmetry, hence with \hbox{$H_{C}(r_{C}) = 0$} it follows that this additional function of $C$ is identical zero and (\ref{eq:tevenfinal}) is found. A non-trivial example, namely ``zgp'' boundary conditions leading to $t(C,r)$ as in (\ref{eq:tzgpfinal}), will be discussed in Sec.~\ref{sec:zgplapse}.

To emphasize the importance of the PDE (\ref{eq:Estadtdr}) governing the foliation of Reissner-Nordstr\"om into maximal slices formulated as level sets of 
\beq
	\sigma = t - t(C,r) = \begin{rm}{const}\end{rm},
\eeq
where $C$ is a parameter depending only on the new time coordinate $\tau$, two further ways of re-deriving this equation will be outlined here.

M.\ Duncan based on a variational principle shows in \cite{Duncan85} that for Reissner-Nordstr\"om maximizing the \hbox{3-volume} of the line element 
\beq
	^{(3)}ds^{2} = \left[ -f(r) \left( \frac{\partial t}{\partial r} \right)^{2} + \frac{1}{f(r)} \right] \:dr^{2} 
			+ r^{2}\:d\Omega^{2}
\eeq 
the Euler-Lagrange equation 
\beq
\label{eq:DYDX}
	\frac{D\left[\frac{r^{2} f(r) \frac{\partial t}{\partial r}}{\sqrt{\frac{1}{f(r)} - f(r)
                                  \left( \frac{\partial t}{\partial r} \right)^{2} }}\right]}{Dr} = 0
\eeq
using the notation \hbox{$\frac{DY}{DX} = \frac{\partial Y}{\partial X} + \frac{\partial t}{\partial X} \frac{\partial Y}{\partial t} + \frac{\partial^{2} t}{\partial X^{2}} \frac{\partial Y}{\partial t_{,X}}$} is obtained.
Eq.~(\ref{eq:DYDX}) is satisfied if the expression within the square brackets is a function of $\tau$ only and one may readily check that (\ref{eq:Estadtdr}) is recovered by setting this function to $C(\tau)$.

Another way followed for Schwarzschild in \cite{Beig98,Beig2000} for maximal and in \cite{Gentle2001} for constant mean curvature slices is to examine the behaviour of the normal 
\beq
	n = n_{t} dt + n_{r} dr
	  = N \nabla \left( t - t(C,r) \right) 
          = N \left( dt - \frac{\partial t}{\partial r} dr \right) ,
\eeq
where with the underlying Reissner-Nordstr\"om line element (\ref{eq:RNradial}) the normalization constant $N$ is fixed by demanding
\beq
\label{eq:Nnormalization}
	n_{\mu} n^{\mu} = N^{2} \left( -\frac{1}{f(r)} + f(r) \left(\frac{\partial t}{\partial r}\right)^{2} \right)
                        = -1.
\eeq
Note that $N$ is also given by
\beq
\label{eq:NBeig}
	N = -n_{\mu} \left( \frac{\partial}{\partial t} \right)^{\mu},
\eeq
where \hbox{$\frac{\partial}{\partial t}$} is the static Killing vector.
Hence $N$ can be viewed as the boost function \cite{Beig2000} of \hbox{$\frac{\partial}{\partial t}$} relative to \hbox{$\sigma = \begin{rm}{const}\end{rm}$}.
With (\ref{eq:Kofn}) it follows that the trace of the extrinsic curvature can be written as
\beq
\label{eq:Kofnresult}
	K = - \nabla_{\mu} n^{\mu} = - \frac{1}{r^{2}} \frac{d\left[ r^{2} n^{r} \right]}{dr}
          = - \frac{1}{r^{2}} \frac{d\left[\frac{r^{2} f(r) \frac{\partial t}{\partial r}}{\sqrt{\frac{1}{f(r)} - f(r)
                                  \left( \frac{\partial t}{\partial r} \right)^{2} }}\right]}{dr}
\eeq
which for maximal slices, \hbox{$K \equiv 0$}, yields again (\ref{eq:Estadtdr}) by the same argument as for (\ref{eq:DYDX}).

One may readily verify that from the normalization (\ref{eq:Nnormalization}) it follows that $N$ coincides with the odd lapse, so
\beq
	N = \pm \frac{\sqrt{p_{C}(r)}}{r^{2}} = \alpha_{odd}^{\pm}.
\eeq
Furthermore, writing \hbox{$\frac{\partial}{\partial t}$} as
\beq
	\left( \frac{\partial}{\partial t} \right)^{\mu} = N n^{\mu} + \xi^{\mu} 
        \ \ \begin{rm}{with}\end{rm}\ \ 
        \xi_{\mu}n^{\mu} = 0
\eeq 
and using (\ref{eq:NBeig}) together with \hbox{$n_{\mu} = - \alpha \nabla_{\mu} \tau$}, one can find
\beq
\label{eq:makenewalpha}
	\alpha = N \frac{dt}{d\tau}.
\eeq
In the context of even boundary conditions this yields \hbox{$\alpha_{even} = \alpha_{odd} \frac{\partial t_{even}}{\partial \tau_{even}}$} as in the first line of (\ref{eq:alphaevenfinal}). 
Here $\alpha_{odd}$ and $\alpha_{even}$ are linearly independent solutions of (\ref{eq:dda=Ra}) as pointed out in detail in \cite{Beig98,Beig2000}.

One should furthermore mention that instead of varying the parameter $C$ with time - which is the choice leading to a foliation, i.e.\ a ``time evolution'', made in the latter references and in our paper - one could also as done in \cite{Gentle2001} consider adding constant ``time translations'' to $t(C,r)$ for fixed $C$. 

\subsection{Initial lapse profile}
\label{subsec:Initialevenlapse}
In order to calculate the initial lapse profile note that for \hbox{$C=0$} the primitive for $K_{C}(r)$ as in (\ref{eq:Kintegral}) is given analytically by
\beq
 	K_{C=0}(r)  =  - \frac{2r(2Mr - Q^{2})\sqrt{f(r)}}
                              {(M^{2} - Q^{2}) (2r^{2}-3Mr+Q^{2})} ,  
\eeq
and in particular in the limit \hbox{$r \to \infty$} by
\beq
\label{eq:Kinfinit}
	K_{C=0}(\infty)  = \lim_{r \to \infty} K_{C=0}(r) =  - \frac{2M}{M^{2} - Q^{2}}.
\eeq
With (\ref{eq:alphaevenfinal}) therefore the initial lapse profile in Schwarzschild coordinates is found to be
\beq
\label{eq:alphaeveninitr}	
	\alpha^{\pm}_{even}(C=0,r) = 1 - \frac{Q^{2}}{Mr}
\eeq
taken symmetrically across the throat with the value \hbox{$\alpha_{even}^{\pm}(C=0,r_{+}) = \sqrt{1-\frac{Q^{2}}{M^{2}}}$} there. In terms of the Schwarzschild isotropic coordinates as in (\ref{eq:rofR}) with the isometry (\ref{eq:isometry}) present this can be written as 
\beq
\label{eq:alphaeveninitR}
	\alpha^{\pm}_{even}(C=0,R) = 1 - \frac{Q^{2}}{M} \frac{4R}{4R^{2} + 4MR + M^{2} - Q^{2}}.
\eeq
Obviously, in the Schwarzschild case \hbox{$\alpha^{\pm}_{even}(C=0,r) = 1$} is a solution of (\ref{eq:dda=Ra}), while with increasing charge the source term \hbox{$4\pi (S - 3\rho)$} in the maximality condition leads to an increasingly collapsed lapse profile.


\section{Puncture Lapse}
\label{sec:zgplapse}

\subsection{Motivation of the zero gradient at the puncture (``zgp'') boundary condition}

As stated in the introduction, the motivation to define and study the ``zgp'' boundary condition derives from puncture evolutions of black holes. 
In any numerical black hole evolution one has to specify how the black hole singularity is treated.
In particular, one can excise the black hole from the domain at or inside the event horizon, thereby creating an inner boundary of the numerical grid, one can use the throat as an inner boundary and impose isometry conditions there, and finally one can work on a punctured $R^{3}$, i.e.\ $R^{3}$ minus one point (the ``puncture'') for every asymptotically flat end inside a black hole.

Under different names, punctures were studied as early as \cite{Misner57,Brill63} for Schwarzschild and axisymmetric black hole data.
The puncture topology for black hole initial data was revived in \cite{Brandt97} as a simple method to construct multiple black hole initial data with linear momentum and spin (see also \cite{Beig94,Beig96}).
As it turns out, even in the general case of orbiting black holes one can analytically separate the coordinate singularity at the puncture and work on $R^{3}$, both for initial data and evolution \cite{Bruegmann97,Alcubierre02a}.

Here we concentrate on evolutions of Schwarzschild in the puncture framework.
The 3-metric of Schwarzschild in isotropic coordinates at \hbox{$t=0$} is 
\beq
	 ^{(3)}ds^2 = (1+\frac{M}{2x})^4 \left( dx^2 + x^2 \: d\Omega^2 \right) ,
\eeq
where the radial grid coordinate $x$ coincides with the isotropic Schwarzschild coordinate $R$ as given in (\ref{eq:Rofr}), but differs at later times depending on the shift and the boundary conditions used during evolution.
The coordinate singularity at the puncture, \hbox{$x=0$}, corresponds to the inner asymptotically flat infinity of the black hole, i.e.\ the puncture corresponds to the left-hand $i^{0}$ in the Carter-Penrose diagrams of Subsec.~\ref{subsec:KSCP}.

For evolutions, one can define conformally rescaled metric and extrinsic curvature variables,
\beq
	g_{ij} = \mbox{}^{0}\Psi^{-4} G_{ij}, \quad 
	k_{ij} = \mbox{}^{0}\Psi^{-4} K_{ij},
\eeq
where \hbox{$^{0}\Psi = 1+\frac{M}{2x}$} is the time independent conformal factor of Schwarzschild in isotropic coordinates.
As argued in \cite{Bruegmann97} and discussed further in \cite{Alcubierre02a}, as a consequence of this rescaling there are no singularities apart from the coordinate singularity in $^{0}\Psi$, in particular the rescaled metric and extrinsic curvature are regular and do not evolve at the puncture.
What concerns us here is the regularity of the maximal slicing condition (\ref{eq:dda=Ra}). 

Using the conformal metric $g_{ij}$, one has to solve equations of the type
\beq
\label{eq:puncturelapse}
	\Delta^{g} \alpha - O(\frac{1}{x}) \partial_k \alpha = 0.
\eeq
As it turns out, standard numerical methods to solve this elliptic equation can find a regular solution for $\alpha$ on $R^3$, 
which has the feature that its gradient vanishes sufficiently rapidly at the puncture such that \hbox{$O(\frac{1}{x}) \partial_k \alpha$} is zero at the puncture, and the lapse collapses to zero near the puncture.
This was noted as an experimental fact in \cite{Anninos95} and analyzed in \cite{Bruegmann97,Alcubierre02a}, but there was no rigorous proof that a regular solution to (\ref{eq:puncturelapse}) exists, and while numerically quite robust, it was not clear that the numerics was able to reliably determine this lapse. 

The proposal for the ``zgp'' lapse is to impose the vanishing of the gradient as a boundary condition on the lapse in coordinates that avoid the coordinate singularity of isotropic coordinates.

\subsection{Derivation of the puncture lapse}
In order to derive boundary conditions for the ``zgp'' lapse, note that at infinity and at the puncture the radial grid coordinate $x$, and also the isotropic Schwarzschild coordinate $R$ at \hbox{$t = 0$} given in (\ref{eq:Rofr}), is related to the Schwarzschild radius through \hbox{$r = x$} for \hbox{$\{r \to \infty,x \to \infty\}$} and \hbox{$r \propto \frac{1}{x}$} for \hbox{$\{r \to \infty, x \to 0\}$}.
As before we demand unit lapse at infinity,
\beq
\label{eq:BCzgp1}
	\lim_{x \to \infty} \alpha^{+}_{zgp} = \lim_{r \to \infty} \alpha^{+}_{zgp} = 1 \ \ \forall \tau_{zgp},
\eeq
such that the slices are labeled by proper time at infinity. 
At the puncture, we impose the condition of vanishing gradient by
\beq
\label{eq:BCzgp2}
	\lim_{x \to 0} \partial_{x} \alpha^{-}_{zgp} \propto \lim_{r \to \infty} r^{2} \partial_{r} \alpha^{-}_{zgp} = 0 
        \ \ \forall \tau_{zgp}
\eeq
making use of \hbox{$\partial_{x} \propto  r^{2} \partial_{r}$} in this limit.

It turns out to be convenient to derive the ``zgp'' lapse by making the ansatz 
\footnote{This was suggested to us by R.\ Beig, whom we would like to thank for very useful conversations and valuable comments.}
\beq
\label{eq:linearAnsatz}
	\alpha^{\pm}_{zgp}(C,r) = \Phi(C) \cdot \alpha^{\pm}_{even}(C,r) + \Psi(C) \cdot \alpha^{\pm}_{odd}(C,r)
\eeq  
for a time-dependent linear combination of the odd and the even lapse.
This form is possible since the maximality condition (\ref{eq:dda=Ra}) is linear in the lapse. Furthermore, with $\alpha_{odd}$ and $\alpha_{even}$ two linearly independent solutions are known as discussed for Schwarzschild in \hbox{Appendix A} of \cite{Beig98} and as is true also for Reissner-Nordstr\"om with \hbox{$0 \leq \left| Q \right| < M$}.
This ansatz reveals useful insights in the properties of odd, even, and ``zgp'' lapse, leading e.g.\ immediately to the conjecture of Subsec.~\ref{subsec:Conjecture} regarding the lapse at the right-hand event horizon at late times.
Also, the smoothness of the lapse (\ref{eq:alphazgpfinal}) and of the height function (\ref{eq:tzgpfinal}) follow trivially from the smoothness shown for odd and even boundary conditions in \cite{Beig98}. 
Equivalently, the puncture lapse can of course be derived starting with the general solution for the lapse, (\ref{eq:generalalpha}), by applying directly the desired boundary conditions (\ref{eq:BCzgp1}) and (\ref{eq:BCzgp2}).

Care has to be taken in demanding these boundary conditions analytically since condition (\ref{eq:BCzgp1}) is imposed on the right-hand side of the throat (denoted by a superscript ``$+$'') while (\ref{eq:BCzgp2}) on the left-hand side (denoted by a ``$-$'').
This, however, will be taken into account in the following as the odd lapse is taken in an antisymmetric manner across the throat using the appropriate plus and minus sign in (\ref{eq:alphaoddfinal}) for the regions extending to infinity and puncture, respectively. The even lapse is symmetric with respect to the throat by construction.

Since both of these two lapses are already one at infinity, the first boundary condition (\ref{eq:BCzgp1}),
\beq
\label{eq:BCzgp1m}
	  \Phi(C) + \Psi(C) = 1 \ \ \forall C,
\eeq
simply yields that the sum of the dimensionless multiplicator functions $\Phi$ and $\Psi$ is one for all times, so one can substitute \hbox{$\Psi(C) = 1-\Phi(C) \ \forall C$}. This (trivial) outer boundary condition already allows to formulate a conjecture regarding the value of the lapse at the right-hand event horizon at late times as stated in Subsec.~\ref{subsec:Conjecture}. 

Continuing in the derivation of the puncture lapse, the gradients of the odd and even lapse at the puncture can be obtained by calculating
\beq
\label{eq:oddevenderiv}
\begin{array}{lcc}
	\lim_{r \to \infty} r^{2} \partial_{r} \alpha^{-}_{odd} 
                                    & = & - M , \\ 
	\lim_{r \to \infty} r^{2} \partial_{r} \alpha^{-}_{even} 
                                    & = & M + \frac{2}{K_{C}(\infty)}.
\end{array}
\eeq
Remembering that the integral $K_{C}(\infty)$ carries units of $\frac{1}{M}$ and diverges for \hbox{$C \rightarrow C_{lim}$}, it turns out that for late times the gradients at the puncture for $\alpha_{odd}$ and $\alpha_{even}$ just differ by their sign. Therefore, in order to produce a vanishing gradient, at late times the average of the two lapse functions has to be taken, i.e.\ 
\beq
\label{eq:latetimeaverage}
	\lim_{C \to C_{lim}} \Phi(C) = \lim_{C \to C_{lim}} \Psi(C) = \frac{1}{2}. 
\eeq
For arbitrary $C$, $0 \leq C \leq C_{lim}$, we use (\ref{eq:oddevenderiv}) with the second boundary condition (\ref{eq:BCzgp2}) to obtain
\beq
\label{eq:BCzgp2m}
	  \Phi(C) \cdot (M + \frac{2}{K_{C}(\infty)}) - (1-\Phi(C)) \cdot M = 0 \ \ \forall C. 
\eeq

Hence from (\ref{eq:BCzgp1m}) and (\ref{eq:BCzgp2m}) the multiplicator functions $\Phi$ and $\Psi$ are found to be given by
\beq
\label{eq:Phi}
	\Phi(C) = 1-\Psi(C)
                = \frac{1}{2}
                  \:\frac{K_{C}(\infty)}{K_{C}(\infty)+\frac{1}{M}} ,
\eeq
where the divergence of $K_{C}(\infty)$ in the limit \hbox{$C \rightarrow C_{lim}$} gives the average of the odd and even lapse for late times as formulated already in (\ref{eq:latetimeaverage}).

The ``zgp'' or puncture lapse now can be written as
\bea
\label{eq:alphazgpfinal}
  			\alpha^{\pm}_{zgp}(C,r) 
                                               & = & 
                         \frac{\sqrt{p_{C}(r)}}{r^{2}}  
			 \frac{dC}{d\tau_{zgp}} 
			 \frac{\partial t_{zgp}}{\partial C} \\
                                            \  & = & 
                         \Phi(C) \cdot \alpha^{\pm}_{even}(C,r) 
                       + \Psi(C) \cdot \alpha^{\pm}_{odd}(C,r) \nonumber \\
                                            \  & = &
                           - \frac{1}{2} \frac{1}{K_{C}(\infty) +\frac{1}{M}}
                           \left( \frac{1}{r - \frac{3}{2}M + \frac{Q^{2}}{2r}} \right. \nonumber \\
                                            \  & \ & \ \ \ 
                           \left. - \frac{\sqrt{p_{C}(r)}}{r^{2}} 
                                            (K_{C}(r) 
                                        \pm (K_{C}(\infty) + \frac{2}{M})) \right) , \nonumber
\eea
making use of the ansatz (\ref{eq:linearAnsatz}).
Similar to the even case, the first line is obtained by passing to Schwarzschild coordinates, which leads to (\ref{eq:Estadtdtau}) and (\ref{eq:Estadtdr}) with the subfix ``zgp'' instead of ``even''. 

We show the ``zgp'' lapse profile in these coordinates at times at infinity \hbox{$\tau_{zgp} = \{0,1M,2M,4M,8M\}$} when we discuss the numerical evolutions, see Fig.~\ref{fig:SSlapseloop}.

The shift and the radial component of the \hbox{3-metric} are again given by
\beq
\label{eq:betazgpfinal}
        \beta_{zgp} (C,r) = \frac{\alpha_{zgp}(C,r)\gamma_{zgp}(C,r)}
                                                      {r^{2}} C
\eeq
and 
\beq 
\label{gammazgpfinal}
	\gamma_{zgp} (C,r) = \frac{1}{f(r) + \frac{C^{2}}{r^{4}}} 
                           = \frac{r^{4}}{p_{C}(r)}.
\eeq

In order to fix the $\tau_{zgp}$-dependence of $C$ one may from (\ref{eq:alphazgpfinal}) by analogy with (\ref{eq:alphaevenfinal}) conjecture that
\bea
\label{eq:zgpdtdC}
	\frac{\partial}{\partial C} t_{zgp}^{\pm}(C,r) 
			  & = & \frac{r^{2}}
                                     {2(r - \frac{3}{2}M + \frac{Q^{2}}{2r})
                                       \sqrt{p_{C}(r)}} \\
                        \ & \ & - \frac{1}{2} \left( K_{C}(r) 
                                          \pm \left(K_{C}(\infty) + \frac{2}{M} \right) \right) \nonumber
\eea
and hence
\beq
\label{eq:zgpdtaudC}
	\frac{d}{dC} \tau_{zgp} (C) = \lim_{r \to \infty} \frac{\partial}{\partial C} t_{zgp}^{+}(C,r) 
				    = - K_{C}(\infty) - \frac{1}{M}.
\eeq
By going back to the differential equation for the maximal lapse, (\ref{eq:differentiallapse}), one can readily verify that (\ref{eq:zgpdtdC}) and hence (\ref{eq:zgpdtaudC}) are correct and therefore the time-dependence is consistently fixed by the height function
\beq
\label{eq:tzgpfinal}
	t_{zgp}^{\pm}(C,r) = H_{C}(r) \pm (H_{C}(\infty) - \frac{C}{M})
\eeq
using (\ref{eq:dCdH}) and (\ref{eq:HisKhalf}). 
Furthermore, time is measured at right-hand spatial infinity by
\beq
\label{eq:tauzgpfinal}
	\tau_{zgp} (C)  = \lim_{r \to \infty} t_{zgp}^{+}(C,r) = 2 H_{C}(\infty) - \frac{C}{M}.
\eeq
Note, however, that the time measured at the puncture approaches the finite value
\beq
\label{eq:lefttime}
	\lim_{C \to C_{lim}} \lim_{r\to \infty} t_{zgp}^{-} (C,r) = \frac{C_{lim}}{M}.
\eeq
We are not aware of a physical meaning of this finite time at the puncture.

Finally, we want to point out that using our result for the ``zgp'' lapse, (\ref{eq:alphazgpfinal}), together with \hbox{$r \propto \frac{1}{x}$} for \hbox{$\{r \to \infty, x \to 0\}$}, one can readily verify that the first and second derivative of the lapse at the puncture both vanish.
So the ``zgp'' lapse is at least ${\cal C}^{2}$ there.
However, when calculating higher derivatives of the lapse, terms of higher order in the relationship between $r$ and $x$ can be of importance, and these terms depend on the chosen shift.
For a vanishing shift the transformation $r(C(\tau),x)$ will be derived in Subsec.~\ref{subsec:Gridcoordinates}.

\subsection{Initial lapse profile}
For the initial lapse profile with (\ref{eq:Kinfinit}) the multiplicator function $\Phi$ as in (\ref{eq:Phi}) for \hbox{$C=0$} is found to be
\beq
\label{eq:Phiinit}
		\Phi(C = 0) = \frac{M^{2}}{M^{2} + Q^{2}}.
\eeq
In particular, for Schwarzschild with \hbox{$^{0}\Phi(C = 0) = 1$} one starts with the even lapse only, whereas for Reissner-Nordstr\"om with increasing charge the balance is shifted more and more to the odd lapse to produce a ``pre-collapsed'' lapse profile. 
 
So the initial puncture lapse profile in terms of the Schwarzschild radial coordinate $r$ using (\ref{eq:alphazgpfinal}) is found to be given on the right-hand and left-hand side of the throat as
\beq
\label{eq:alphazgpinit}
	\alpha^{\pm}_{zgp}(C=0,r) = \frac{M^{2}r - Q^{2} M \pm Q^{2}r\sqrt{f(r)}}
                                                  {(M^{2} + Q^{2})r}.
\eeq
Applying the transformation (\ref{eq:rofR}) one can derive this profile also for the Schwarzschild isotropic coordinate $R$,
\beq
\label{eq:alphazgpinitR}
	\alpha^{\pm}_{zgp}(C=0,R) = 1 - \frac{2Q^{2}}{M^{2}+Q^{2}}\frac{4R+M^{2}-Q^{2}}{4R^{2}+4MR+M^{2}-Q^{2}}.
\eeq
Note that for the puncture lapse the isometry (\ref{eq:isometry}) is no longer present. 

\subsection{Late time limit for the lapse at the right-hand event horizon}
\label{subsec:Conjecture}
Based on the ansatz (\ref{eq:linearAnsatz}) we want to state a conjecture for the lapse at the right-hand event horizon at late times that is of some importance for black hole excision. 
Although one can use the apparent horizon as the location for the excision boundary, finding the apparent horizon is usually time consuming, and a simpler prescription for some surface that coincides or is located inside the event horizon can serve the same purpose for excision. 
For example, one sometimes can define the excision surface by a surface of constant lapse. 
In numerical simulations with maximal slicing, a lapse between $0.3$ \cite{Seidel2002} and $0.34$ \cite{Walker98} turned out to mark the location of the right-hand event horizon.
A value in this range can also be inferred from some graphs in the literature showing the lapse as a function of time, c.f.\ Fig.~3(a) in \cite{Smarr78b}, Fig.~6 in \cite{Bernstein94} or Fig.~13 in \cite{Brewin2001}. 

For very general boundary conditions we can state the following conjecture:
\newline
\begin{it}
For maximal slicing of the non-extremal extended Reissner-Nordstr\"om spacetime with the throat of the Einstein-Rosen bridge coinciding initially with the event horizon and with boundary conditions imposed on the lapse other than the odd ones, the lapse at late times at the right-hand event horizon is found to approach the value \hbox{$\alpha |_{r_{+}} = \frac{C_{lim}}{r_{+}^{2}}$} asymptotically. 
In particular, for the Schwarzschild spacetime with \hbox{$^{0}C_{lim} = \frac{3}{4}\sqrt{3} M^{2}$} and \hbox{$^{0}r_{+} = 2M$} this value turns out to be \hbox{$^{0}\alpha |_{r_{+}} = \frac{3}{16} \sqrt{3} \approx 0.3248$}. 
\end{it}

Without attempting a complete proof, we can argue as follows.
Since for fixed time, i.e.\ for fixed slice label $C$, the lapse arises from the maximality condition (\ref{eq:dda=Ra}), i.e.\ a second order linear ODE with smooth coefficients, the theory of ODE states that the set of solutions forms a 2-dimensional real linear space, c.f.\ Theorem I in \S 15 of \cite{Walter98}. 
As a consequence, every lapse using the superposition principle can be constructed as a (time-dependent) linear combination of the odd and even lapse, which are evidently linearly independent for \hbox{$0 \leq \left| Q \right| < M$}. 
From (\ref{eq:alphaoddfinal}) and (\ref{eq:alphaevenfinal}) evaluated for \hbox{$C \to C_{lim}$}, one can see that odd and even lapse at the right-hand event horizon in this limit are given by
\beq
	\lim_{C \to C_{lim}} \alpha_{odd}^{+} \mid_{r_{+}} 
      = \lim_{C \to C_{lim}} \alpha_{even}^{+} \mid_{r_{+}} 
      = \frac{C_{lim}}{r_{+}^{2}}.
\eeq
Furthermore, any other lapse constructed by a linear combination with coefficients adding up to one in order to have unit lapse at infinity has the same limiting value. 
Finally, we argue that as the slices approach with \hbox{$r_{C} \to r_{C_{lim}} > 0$} a limiting surface and hence avoid the singularity \cite{Eardley79}, the limit \hbox{$C \to C_{lim}$} considered so far carries over to the limit of infinite time, \hbox{$\tau \to \infty$}.
Although reasonable, it is not yet clear to us whether the last step holds without exception.


\section{Comparison of the analytical odd, even, and zgp maximal slices}
\label{sec:compareanalytic}

\subsection{Embedding diagrams}
\label{subsec:embedding}
Embedding diagrams are useful tools for the visualization of geometric properties of slicings.
Following Appendix F of \cite{Bernstein93} and Sec.~IV of \cite{Brill79} in suppressing one of the angular coordinates, by embedding a 2-dimensional Riemannian hypersurface into flat 3D space the intrinsic geometry of this surface is preserved.
The embedding of the Reissner-Nordstr\"om \hbox{$\{r,\phi\}$} line element, i.e.\ the equatorial hypersurface \hbox{$\theta = \frac{\pi}{2}$} corresponding to fixed time at infinity $\tau$, in Euclidean space written in cylindrical coordinates \hbox{$\{z, r=\sqrt{x^{2}+y^{2}},\phi\}$} is obtained by demanding
\bea
	^{(3)}ds^{2} & = & dz^{2} + dr^{2} + r^{2}\:d\phi^{2} \nonumber \\
               	     & = & \left[- f(r) \left(\frac{\partial t}{\partial r}\right)^{2} + \frac{1}{f(r)}\right]\:dr^{2} 
                           + r^{2}\:d\phi^{2} \nonumber \\
                     & = & \frac{r^{4}}{p_{C}(r)}\:dr^{2} + r^{2}\:d\phi^{2}.
\eea
For fixed $C(\tau)$ this results in the differential equation 
\beq
\label{eq:dzdr}
	\frac{dz}{dr} \Bigg|_{C = \begin{rm}{const}\end{rm}} = \pm \sqrt{\frac{r^{4}}{p_{C}(r)} - 1}
\eeq
for the embedding function $z(C,r)$. 
For \hbox{$0 \leq \left| Q \right| < M$} and \hbox{$r \geq r_{C}$} its solution is found by integration,
\beq
\label{eq:zofr}
	z(C,r) = \pm \int_{r_{C}}^{r} 
		              \sqrt{\frac{y^{4}}{p_{C}(y)} - 1}\:dy,
\eeq
where without loss of generality symmetry with respect to the \hbox{$z=0$} plane has been chosen.
In particular for the initial Schwarzschild slice, i.e.\ for \hbox{$Q = C = 0$}, the solution \cite{Bernstein93} is found to be
\beq
	^{0}z(C=0,r) = 2\sqrt{2} \sqrt{(r-2M)M}.
\eeq
The embedding is a parabola of revolution referred to as Einstein-Rosen bridge \cite{Einstein35}. 
Here in both ``universes'' the 3-geometry becomes Euclidean far from the throat, which is the minimal 2-sphere of the bridge. 

Note, however, that Einstein's field equations fix only the local geometry of the spacetime but not its topology. A geometrically identical - but topologically different - embedding could be found for example by identifying both sheets in order to obtain the ``throat of a wormhole'' in the sense of Misner and Wheeler \cite{Misner57}.
Here in one flat space two distant regions are connected in the limit when the separation of the wormhole mouths is large compared to the circumference of the throat, c.f.\ Fig.~31.5 in \cite{Misner73}.

\begin{figure}[!ht]
	\noindent
	\epsfxsize=100mm \epsfysize=150mm \epsfbox{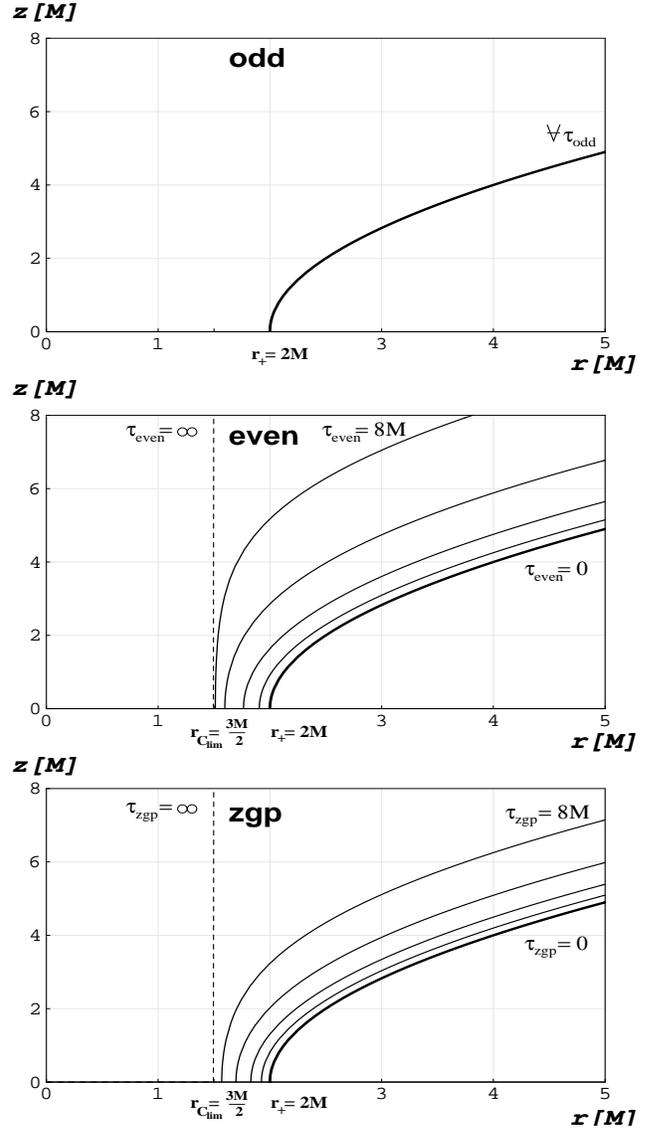} 
	\caption{In the embedding diagrams for maximally slicing the Schwarzschild metric using odd, even, and ``zgp'' boundary conditions, the Einstein-Rosen bridges are obtained by reflecting the curves corresponding to \hbox{$\tau=\{0,1M,2M,4M,8M\}$} displayed from the bottom up at \hbox{$z=0$} and rotating them around the z-axis. They start with the throat at \hbox{$^{0}r_{+} = 2M$} and degenerate for \hbox{$\tau \to \infty$} in the even and ``zgp'' case - which just differ by their time labeling - to an infinitely long cylinder with radius \hbox{$^{0}r_{C_{lim}} = \frac{3M}{2}$}.}
	\label{fig:SSembedding}
\end{figure}

For the general \hbox{$0 \leq \left| Q \right| < M$}, \hbox{$0 \leq C \leq C_{lim}$} case we evaluated the integral (\ref{eq:zofr}) numerically using for the relationship $C(\tau)$ equations (\ref{eq:tauevenfinal}) and (\ref{eq:tauzgpfinal}) for the even and ``zgp'' boundary conditions, respectively. 
The embeddings of these two cases therefore only differ by their time labeling and are less informative than their corresponding spacetime diagrams shown in Subsec.~\ref{subsec:KSCP}. 
In particular, information about any underlying symmetry of the slices - like the (anti-)symmetry with respect to the throat for (odd) even boundary conditions and the more complicated behaviour for ``zgp'' conditions - has been lost.
The results are shown for Schwarzschild in Fig.~\ref{fig:SSembedding}, which should be compared for the even boundary conditions with Fig.~4.18 in \cite{Bernstein93} or Fig.~2 in \cite{Estabrook73}.
Reflecting the curves at the \hbox{$z=0$} plane and rotating them around the $z$-axis, a set of throats is obtained starting with a parabola of revolution with the minimal 2-sphere initially at the event horizon \hbox{$^{0}r_{+} = 2M$} and degenerating at late times to an infinitely long cylinder with radius \hbox{$r =$ $^{0}r_{C_{lim}} = \frac{3M}{2}$} as the slices approach the maximal hypersurface with this radius asymptotically. 
The plots for charge in the range \hbox{$0 < \left| Q \right| < M$}, which are not shown here, are qualitatively very similar.

\subsection{Kruskal-Szekeres and Carter-Penrose diagrams}
\label{subsec:KSCP}
Spacetime diagrams such as the Kruskal-Szekeres \cite{Kruskal60} and Carter-Penrose \cite{Carter66a,Carter66b} diagrams provide a convenient way of visualizing spacetime properties like causality, see e.g.\ \cite{Misner73,Hawking73a}.
In this paper we want to limit ourselves to the discussion of the Schwarzschild spacetime, the results for the non-extremal Reissner-Nordstr\"om spacetime can be found in \cite{mythesis}.

For the Schwarzschild metric (\ref{eq:RNradial}), radial null geodesics motivate for \hbox{$r \in [^{0}r_{+}=2M, \infty[$} the introduction of the Regge-Wheeler radial coordinate
\beq
	r_{*} = \int \frac{dr}{1 - \frac{2M}{r}}
              = r + 2M \ln{\left| \frac{r-2M}{2M} \right|},\ r_{*} \in ]-\infty,\infty[.
\eeq
As \hbox{$d(t\pm r_{*}) = 0$} on radial null geodesics, ingoing Eddington-Finkelstein coordinates \hbox{$\{v,r,\theta,\phi\}$} with
\beq
\label{eq:inEF}
        v = t+r_{*},\ v \in ]-\infty,\infty[
\eeq  
leading to
\hbox{$ds^{2} = -(1-\frac{2M}{r})\:dv^{2} + 2\:dvdr + r^{2}\:d\Omega^{2}$}
can be defined as in \cite{Finkelstein58}. 
Similarly, outgoing Eddington-Finkelstein coordinates \hbox{$\{u,r,\theta,\phi\}$} given by
\beq
\label{eq:outEF}
        u = t-r_{*},\ u \in ]-\infty,\infty[
\eeq 
lead to 
\hbox{$ds^{2} = -(1-\frac{2M}{r})\:du^{2} - 2\:dudr + r^{2}\:d\Omega^{2}$}.
These line elements can now be analytically extended to \hbox{$r>0$}, and the corresponding Eddington-Finkelstein diagrams are shown e.g.\ in \cite{Hawking73a}. 
In these coordinates the metric components are non-singular at \hbox{$^{0}r_{+}=2M$}.
However, it should be mentioned that the regions \hbox{$r<2M$} in ingoing and outgoing Eddington-Finkel\-stein coordinates do not coincide, since one can readily verify that for all timelike or null worldlines from \hbox{$ds^{2}\leq0$} it follows that \hbox{$2dvdr \leq 0$}, while \hbox{$2dudr \geq 0$}.
Therefore, the nature of the singularity at \hbox{$r=0$} is different: 
An analysis in ingoing Eddington-Finkelstein coordinates shows that no signal can escape back to infinity once it passed the event horizon.
The time reverse of a black hole, a white hole, is found by performing a similar analysis in outgoing Eddington-Finkelstein coordinates.  

With the exterior region covered by both in- and outgoing Eddington-Finkelstein coordinates, one can write the Schwarzschild metric as
\bea
\label{eq:Kruskalds}
	ds^{2} & = & -(1-\frac{2M}{r})\:dudv + r^{2}\:d\Omega^{2} \nonumber \\
               & = & - \frac{32M^{3}}{r}\:\epower^{-\frac{r}{2M}}\:dUdV 
                     + r^{2}\:d\Omega^{2}
\eea
by introducing for \hbox{$r>2M$} the Kruskal-Szekeres coordinates \cite{Kruskal60}
\beq
	U = -\epower^{-\frac{u}{4M}} < 0 
        \ \ \begin{rm}{and}\end{rm} \ \ 
        V = +\epower^{+\frac{v}{4M}} > 0,
\eeq
which are analytically extendable to \hbox{$U>0$} and \hbox{$V<0$}. 
The Schwarzschild radial coordinate r is now given implicitly as a function of U and V by
\beq
	U \cdot V  =  -\epower^{\frac{r_{*}}{2M}}
                =  (1-\frac{r}{2M})\epower^{\frac{r}{2M}}. 
\eeq
Note that the singularity at \hbox{$r=0$} corresponds to \hbox{$U \cdot V = 1$} and the event horizon at \hbox{$^{0}r_{+} = 2M$} to \hbox{$U \cdot V = 0$}, i.e.\ to either \hbox{$U = 0$} or \hbox{$V = 0$}. 
It is convenient to introduce new time and space coordinates by
\beq
\label{eq:TXofUV}
	T = \frac{1}{2} (V+U) 
	\ \ \begin{rm}{and}\end{rm} \ \ 
        X = \frac{1}{2} (V-U)
\eeq
in order to produce Kruskal-Szekeres diagrams.
Note that with $t$ expressed in terms of $T$ and $X$ by
\beq
	t = 4M 
            \left\{ 
              \begin{array}{c}
                 \begin{rm} {arctanh} \end{rm}\frac{T}{X} \\
                 \begin{rm} {arctanh} \end{rm}\frac{X}{T} 
              \end{array} 
            \right.
            \ \ \begin{rm}{in\ the\ regions}\end{rm} \ \ \
	    \left\{ 
              \begin{array}{c}
                 \begin{rm}{I,I'}\end{rm} \\
                 \begin{rm}{II,II'}\end{rm}
              \end{array} 
            \right. ,
\eeq
the lines of constant t correspond to straight lines through the origin, whereas curves of constant $r$ are hyperbolae with asymptotes \hbox{$T = \pm X$}.

\begin{figure}[!hb]
	\noindent
	\epsfxsize=67.5mm \epsfysize=195mm \epsfbox{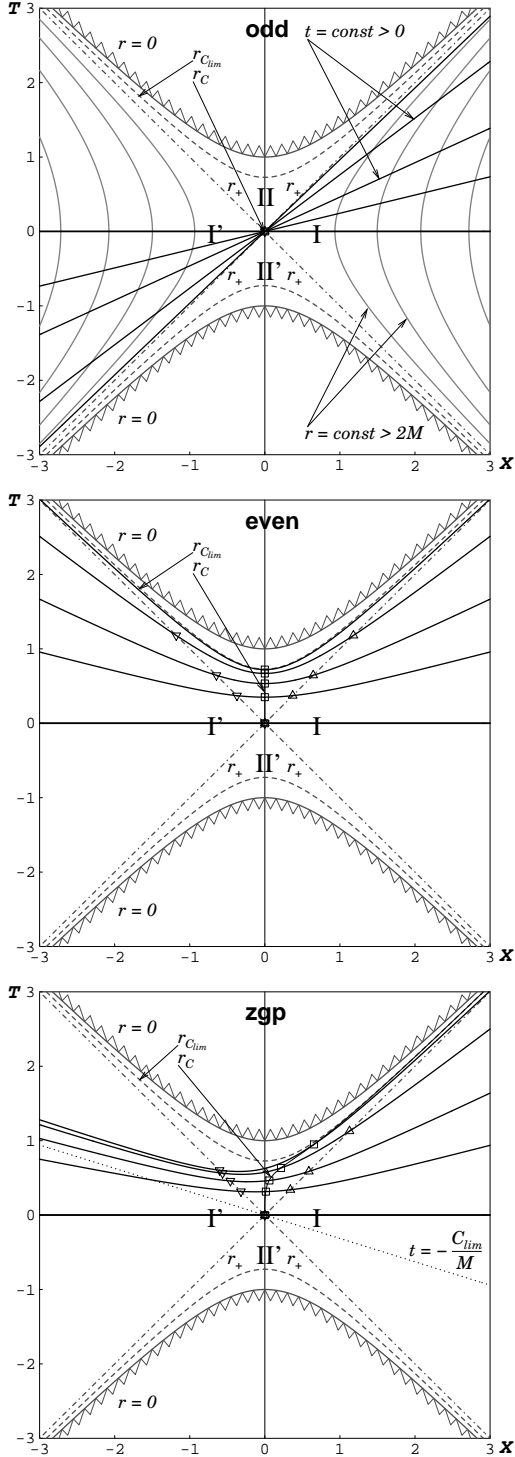}
	\caption{The Kruskal-Szekeres diagrams for the extended Schwarzschild spacetime show the maximal slices for time at infinity \hbox{$\tau = \{0,1M,2M,4M,8M\}$} for the odd, even, and ``zgp'' boundary conditions.  
Note that the puncture corresponds to the spatial infinity of region~I' located to the left of the plot.}
	\label{fig:SSKruskal}
\end{figure}

In Fig.~\ref{fig:SSKruskal}, we show Kruskal-Szekeres diagrams for odd, even, and ``zgp'' lapse. (Note the difference to the hand-drawn Fig.~1 in \cite{Alcubierre02a}.) 
Since the in- and outgoing radial null geodesics are given by lines of constant U and V, the light cones are unit light cones appearing at 45 degree. 
One can infer from the Kruskal-Szekeres diagram that the Schwarzschild geometry consists of four regions separated by the (dot-dashed plotted) event horizon at \hbox{$^{0}r_{+} = 2M$}, namely the regions I and I', two identical but distinct asymptotically flat universes, where the isometry \hbox{$^{0}R \longleftrightarrow \frac{M^{2}}{4 ^{0}R}$} corresponds to the mapping \hbox{$(T,X) \longleftrightarrow (-T,-X)$}, and the regions II and II', two identical but time-reversed regions in which physical singularities (a black hole and a white hole) are present.

As discussed in Sec.~\ref{sec:oddlapse}, the maximal slices with odd boundary conditions correspond to surfaces of constant Schwarzschild time coordinate $t$ in the outer regions I and I' since $\frac{\partial}{\partial t}$ generates a Killing vector field there \cite{Beig98,Townsend97}. 
In particular, in the upper plot of Fig.~\ref{fig:SSKruskal} the solid black straight lines in regions I and I' are (starting with the horizontal time-symmetric \hbox{$t=0$} slice) the spacelike maximal hypersurfaces \hbox{$t=\{0,1M,2M,4M,8M\}$}, whereas the gray timelike lines correspond to \hbox{$r=\{2.5M, 3M, 3.5M, 4M\}$}. 
In regions II and II' the spacelike limiting maximal slice \hbox{$r$ $=$ $^{0}r_{C_{lim}} = \frac{3M}{2}$} is plotted as a dashed line, the singularity \hbox{$r=0$} by a zigzag line. 

In the even case, plotting the height function $t_{even}(C(\tau_{even}),r)$ as in (\ref{eq:tevenfinal}) for the previously stated times at infinity, an observer moves forward in Schwarzschild time $t$ in region~I and backward in region~I' in a symmetric manner in order to reach the limiting maximal slice, c.f.\ the plot in the middle of Fig.~\ref{fig:SSKruskal}, and e.g.\ Fig.~1 in \cite{Estabrook73} or Fig.~3 in both \cite{Bernstein89} and \cite{Geyer95}. 
However, in terms of the new time coordinate $\tau_{even}$, time runs forward (due to the always positive lapse) equally fast (due to the symmetry) in both regions so that the asymptotic values for $t$ at puncture and infinity differ by the amount $2 \tau_{even}$. 
Due to the symmetry with respect to the T-axis, the throat $^{0}r_{C}$ (plotted as an unfilled box) moving from \hbox{$^{0}r_{+}=2M$} to \hbox{$^{0}r_{C_{lim}}=\frac{3M}{2}$} remains on this axis and the slices penetrate into the left-hand and right-hand event horizon (denoted by downward and upward pointing triangles) symmetrically. 

Demanding for ``zgp'' boundary conditions neither symmetry nor antisymmetry, the slices are ``lop-sided'' as can be seen by plotting $t_{zgp}(C(\tau_{zgp}),r)$, (\ref{eq:tzgpfinal}), in the graph on the bottom of Fig.~\ref{fig:SSKruskal}.
In particular, one can observe that as a function of time the throat $^{0}r_{C}$ moves ``to the right'' along the shown curve corresponding to \hbox{$t = H_{C}(\infty) - \frac{C}{M}$}. 
From (\ref{eq:tzgpfinal}) one can infer that as in region~I at infinity the time measured as \hbox{$\tau_{zgp}(C) = \lim_{r \to \infty} t_{zgp}^{+} (C,r)$} goes to infinity in the limit \hbox{$C$ $\to$ $^{0}C_{lim}$}, as stated in (\ref{eq:lefttime}) in region~I' at the puncture the finite time \hbox{$\frac{^{0}C_{lim}}{M} = \frac{3}{4}\sqrt{3}M \approx 1.2990M$} is found. 
Taking care of the sign as for even boundary conditions, the maximal hypersurfaces approach in region~I' the line \hbox{$t = - \frac{^{0}C_{lim}}{M}$} asymptotically in the limit \hbox{$r \to \infty$} with \hbox{$C$ $\to$ $^{0}C_{lim}$}.

\begin{figure}[!hb]
	\noindent
	\epsfxsize=90mm \epsfysize=150mm \epsfbox{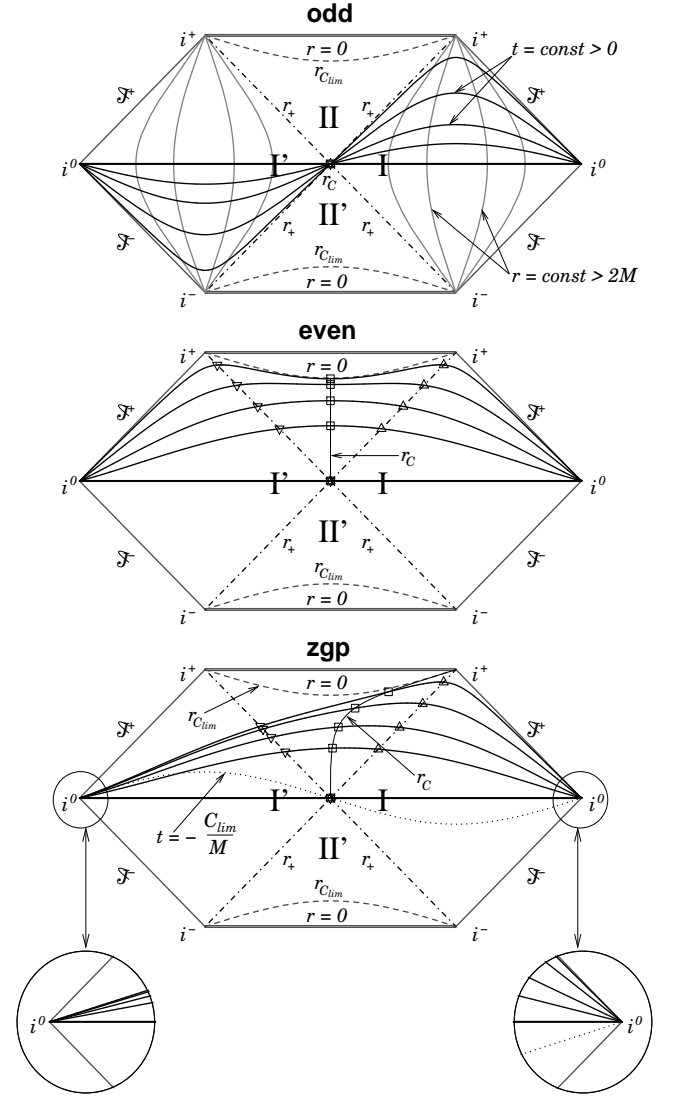}
	\caption{Carter-Penrose diagrams are obtained by compactifying the Kruskal-Szekeres diagrams of Fig.~\ref{fig:SSKruskal} as stated in the text.
Note the behaviour of the slices at spatial infinity, the movement of the throat $r_{C}$ and of the left-hand and right-hand event horizon, which are again plotted as unfilled box, downward and upward pointing triangles, respectively.
}
	\label{fig:SSCP}
\end{figure}

In Fig.~\ref{fig:SSCP}, we show Carter-Penrose diagrams corresponding to Fig.~\ref{fig:SSKruskal}. 
These are obtained by a conformal compactification of the form \hbox{$d\tilde{s}^{2} = \Omega^{2} (t,x^{i}) \cdot ds^{2}$} which - leaving the underlying causal structure unchanged - maps points at infinity in the original metric $ds^{2}$ to a finite affine parameter in the compactified metric $d\tilde{s}^{2}$.
See \cite{Hawking73a} for further technical conditions on the spacetime and on the (not unique) choice of $\Omega$ to guarantee that this construction will work. 
As shown in this reference, on the conformal boundary one can identify past and future null infinity ${\cal J}^{\mp}$, past and future timelike infinity $i^{\mp}$ and spacelike infinity $i^{0}$, i.e.\ the beginning and end of null, timelike and spacelike geodesics. 

\pagebreak

For the Schwarzschild metric, compactification can be carried out \cite{Townsend97} by introducing as in (\ref{eq:TXofUV}) new time and space coordinates $\widetilde{T}$ and $\widetilde{X}$ based on
\beq
\label{eq:compact}
	\widetilde{U} = \arctan{U} 
	\ \ \begin{rm}{and}\end{rm} \ \ 
        \widetilde{V} = \arctan{V},
\eeq
where $\widetilde{U}$ and $\widetilde{V}$ are in the range \hbox{$-\frac{\pi}{2} \leq \widetilde{U},\widetilde{V} \leq \frac{\pi}{2}$}.

It is worth pointing out that as radial null geodesics are again unit light cones the maximal slices are obviously spacelike. 
Furthermore, at the right-hand $i^{0}$, with the lapse being one at infinity for all three boundary conditions, essentially \hbox{$t = $ const} slices are found, whereas in particular for ``zgp'' boundary conditions at the puncture, i.e.\ the left-hand $i^{0}$, the line \hbox{$t = -\frac{^{0}C_{lim}}{M}$} is approached at late times. 
One may readily verify that for the compactification carried out here these slices arrive at $i^{0}$ with a slope given by
\bea
	\left| \frac{d\widetilde{T}}{d\widetilde{X}} \right|_{i^{0}} 
		& = & \begin{rm}{tanh}\end{rm} \frac{|t|}{4M}.
\eea
In our case, starting with the horizontal time-symmetric hypersurface $t = 0$ the slices for all three boundary conditions fan out to hug ${\cal J}^{+}$ asymptotically, while for ``zgp'' boundary conditions at the puncture a finite angle given by \hbox{$\begin{rm}{arctan}\end{rm} \left[ \begin{rm}{tanh}\end{rm} \frac{C_{lim}}{4M^{2}} \right] \approx 17.4^{\circ}$} is found in the limit of late times. 


\section{Comparison with numerical simulations}
\label{sec:numericallapse}

\subsection{Numerical method}

Maximal slicing has been studied numerically for the evolution of a single Schwarzschild black hole already in \cite{Estabrook73}, and was later on implemented and used as a standard test case in 1D- \cite{Bernstein89}, 2D- \cite{Bernstein94} and 3D-codes \cite{Anninos95,Alcubierre02a} based on a Cartesian grid and by using a smooth lattice method \cite{Brewin2001}.

For comparison with the analytical result for the puncture lapse, we have performed simulations with the Cactus code \cite{Cactusweb}.
We evolved a single Schwarzschild puncture with maximal slicing and vanishing shift with the Baumgarte-Shapiro-Shibata-Nakamura (BSSN) system of evolution equations, see \cite{Alcubierre02a} for details. 
The maximal slicing lapse is obtained by solving the elliptic equation (\ref{eq:dda=Ra}) using the multigrid solver BAM\_Elliptic \cite{Bruegmann97a,Bruegmann99}, without imposing any additional boundary condition at the puncture.

\begin{figure}[!ht]
	\noindent
	\epsfxsize=90mm \epsfysize=150mm \epsfbox{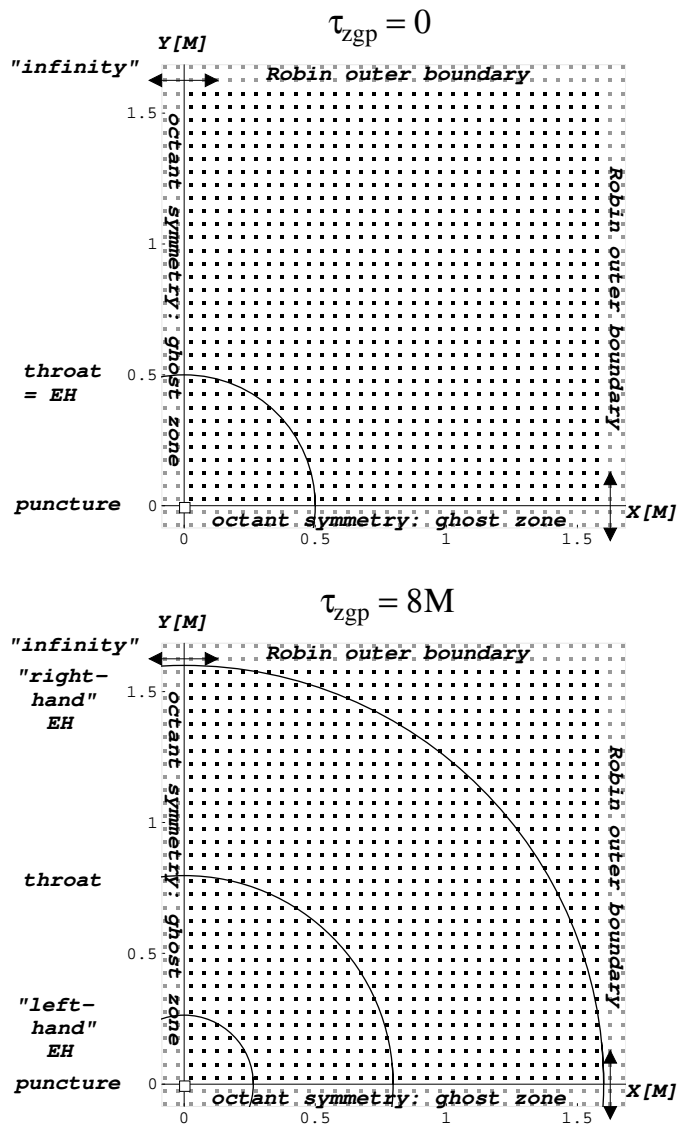}
	\caption{
The Cartesian grid used in numerical solutions is shown (suppressing the Z-direction) for $32^{3}$ grid points and a resolution of $\triangle X = \triangle Y = \triangle Z = 0.05M$, which places the outer boundary at $X_{ob} \approx 1.6M$.
The domain is one octant with appropriate boundary conditions for spherical symmetry imposed at the planes $X=0$, $Y=0$, and $Z=0$.
Note that the puncture is staggered at the origin.
The numerically found locations of the throat $x_{C}$ and the left-hand and right-hand event horizon, $x_{C}^{-}$ and $x_{C}^{+}$, are superimposed as dark solid lines for the times $\tau_{zgp} = 0$ and $\tau_{zgp} = 8M$. 
        }
	\label{fig:Grid}
\end{figure}

In Fig.~\ref{fig:Grid}, we show an example for the numerical grid.
The runs are carried out in 3D on an octant of $R^3$, which is possible for spherically symmetric systems. 
The puncture is located at the origin, and we choose a staggered Cartesian grid with uniform grid spacing such that the puncture is not part of the grid but rather is located half-way between grid points in all three directions.

The outer boundary condition for the lapse is implemented numerically at the edge of the finite numerical grid by employing a Robin condition \cite{York82} as described in \cite{Alcubierre2000}. 
In order to study errors introduced by the Robin boundary condition implemented at the edge of the grid, runs have been carried out using a uniform spacing of \hbox{$\triangle X = 0.05M$} and (multiplied by a courant factor of $\frac{1}{4}$) time steps of \hbox{$\triangle \tau_{zgp} = \frac{1}{4} \triangle X = 0.0125M$} for a series of $32^{3}$, $64^{3}$ and $128^{3}$ grid points to place the outer boundary at \hbox{$X_{ob} \approx 1.6M$}, $3.2M$ and $6.4M$, respectively.
Similarly, we checked for second order convergence using different numbers of grid points for the outer boundary close in.
These numerical tests are described in \cite{mythesis}.
For the resolutions considered here, there are limitations in the numerical convergence near the puncture, and it is known that higher resolution is required to see clean convergence near the puncture, see \cite{Alcubierre02a} for algebraic slicings. 
For the present purpose of comparison with the analytical results, the goal was only to establish that these runs are approximately in the convergent regime.
All the results in the remainder of this section are for \hbox{$\triangle X = 0.05M$}, $128^{3}$ grid points, and \hbox{$X_{ob} \approx 6.4M$}.

We will show that the numerical evolution obtained in this way agrees nicely (within a certain error) with the analytical result for the puncture lapse.
In order to carry out this comparison, we have to establish an explicit mapping between the coordinates used in the numerical simulation and our analytical study.

\subsection{Schwarzschild radial coordinate}
One has to remember that for 3D Cartesian coordinates \hbox{$\left\{ X,Y,Z\right\}$} the system has no spherical symmetry from the point of view of the code.
However, looking at radial output the radial coordinate $x$ defined as the Euclidean distance from the puncture is readily obtained. 
Because of the spherical symmetry the expression $r^{2}\:d\Omega^{2}$ for the angular part of the metric is also expected in numerical simulations \footnote{Furthermore, we observed that $r(\tau_{zgp},x)$ obtained independently as square root of the prefactor of the angular metric part and via the Cactus thorn ``Extract'' coincides up to numerical error.
The latter thorn is normally used for gravitational wave extraction for observers at some distance from the source that coincides for a Schwarzschild black hole with the Schwarzschild radial coordinate $r$ calculated from an expansion of surface integrals of metric components.}. 
Hence the Schwarzschild radial coordinate can be obtained by calculating 
\beq
\label{eq:getr}
	r(\tau,x) = \Psi^{2}(x) \sqrt{g_{\theta\theta}(\tau,x)} 
	          = \Psi^{2}(x) \sqrt{\sin^{2}{\theta} g_{\phi\phi}(\tau,x)}.
\eeq

\begin{figure}[!ht]
	\noindent
	\epsfxsize=85mm \epsfysize=160mm \epsfbox{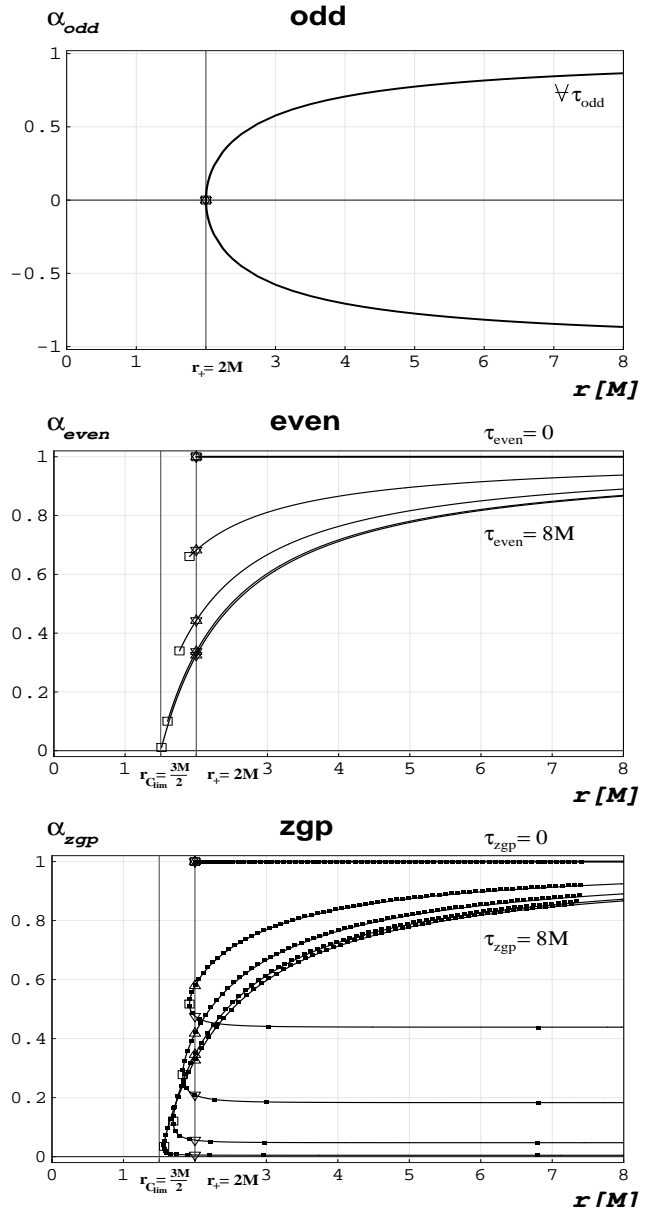} 
	\caption{
Shown are the odd, even, and ``zgp'' lapse versus the Schwarzschild radial coordinate $r$ for times \hbox{$\tau = \{0,1M,2M,4M,8M\}$} measured at infinity. 
The solid lines show the analytic results, while the numerical results for the ``zgp'' lapse are plotted as solid boxes. The numerical data points lie virtually on top of the analytical result.
Note that in the odd case the time-independent lapse is negative on the left-hand side of the throat (denoted by an unfilled box) coinciding with the event horizon, whereas for the even case the collapsing lapse is symmetric in the sense that in particular left-hand and right-hand event horizon (shown as downward and upward pointing triangles) and also infinity and puncture coincide. These symmetries are lost for the ``zgp'' lapse with the lapse remaining one at infinity and collapsing at the puncture.}
	\label{fig:SSlapseloop}
\end{figure}

Fig.~\ref{fig:SSlapseloop} shows the odd, even, and ``zgp'' lapse profiles 
for times \hbox{$\tau = \{0,1M,2M,4M,8M\}$} measured at infinity. 
The solid lines show the analytic results based on the evaluation of (\ref{eq:alphaoddfinal}), (\ref{eq:alphaevenfinal}) and (\ref{eq:alphazgpfinal}).
For the ``zgp'' boundary condition, data points obtained from the largest simulation are plotted as boxes in Fig.~\ref{fig:SSlapseloop} and lie virtually on top of the analytical results.

Evaluating the analytic integrals is a non-trivial task due to the late time divergence of the involved integrals $H_{C}(r)$ and $K_{C}(r)$ as discussed in more detail in \cite{mythesis} based on \cite{Thornburg93}.
The value of the lapse at the throat $^{0}r_{C}$ is marked on these slices by an unfilled box and similarly for the left-hand and right-hand event horizon corresponding to \hbox{$^{0}r_{+} = 2M$} by downward and upward pointing triangles. 
Being motivated geometrically these ``markers'' essentially allow one to sketch for a certain time at infinity the corresponding lapse profile and are hence suited for a study of their late time behaviour which will be carried out in a later paper \cite{mypaper2}. 
Note, for example, that for the collapsing lapse profiles obtained for even and ``zgp'' boundary conditions the radial coordinate of the throat approaches \hbox{$^{0}r_{C_{lim}} = \frac{3M}{2}$} and that the conjecture of Subsec.~\ref{subsec:Conjecture} regarding the lapse at the right-hand event horizon holds. 
\vspace*{0.25cm}

\subsection{Radial isotropic grid coordinates}
\label{subsec:Gridcoordinates}
From the analytical point of view the task of finding the foliation of the extended Reissner-Nordstr\"om metric using maximal slices with ``zgp'' boundary conditions has been solved already in Sec.~\ref{sec:zgplapse} in the sense that in a well-suited coordinate system, namely the radial gauge, (\ref{eq:4mradial}), the maximal slices arising as level sets \hbox{$\tau_{zgp} = \begin{rm}{const}\end{rm}$} of the height function \hbox{$t_{zgp}(C(\tau_{zgp}),r)$}, (\ref{eq:tzgpfinal}), have been found.
Keeping in mind the 3+1 split, the foliations shown previously in Kruskal-Szekeres and Carter-Penrose diagrams are uniquely determined by the lapse function $\alpha_{zgp}^{\pm}(C(\tau_{zgp}),r)$, (\ref{eq:alphazgpfinal}), telling one how to progress from slice $\Sigma_{\tau}$ to the consecutive one $\Sigma_{\tau + \delta \tau}$. 
Since for puncture data in Cactus neither the Schwarzschild coordinates, \hbox{$\{t,r,\theta,\phi\}$}, nor the previously introduced coordinates \hbox{$\{\tau,r,\theta,\phi\}$}, but depending on the shift different ``grid coordinates'', \hbox{$\{\tau,x,\theta,\phi\}$}, are used, it is for a comparison with numerical results important to find the coordinate transformations relating these gauges.
Above all we are now interested in using the freedom of choosing different coordinates ($x^{i}$) on these slices in order to construct coordinates such that the 4-metric coincides at all times with output from a numerical evolution of black hole puncture data and hence to prove by construction the existence of an underlying analytical solution. 
The radial coordinates of this puncture data line element we shall call ``isotropic grid coordinates'' in the case of a vanishing shift and ``isothermal grid coordinates'' for a non-trivial shift to emphasize the connection both to the isotropic and isothermal gauge and to the numerical grid.
Unless stated otherwise, here and in the following the expressions are valid for arbitrary boundary conditions imposed on the lapse.
However, one should note that for odd boundary conditions the time-independent isotropic coordinates of Sec.~\ref{sec:oddlapse} are recovered, which are not horizon-penetrating.  
In this paper we concentrate on the case of vanishing shift, \hbox{$B \equiv 0$}, where $B$ denotes the shift for isotropic coordinates to distinguish it from the shift $\beta$ in radial gauge.
The more general case of a shift \hbox{$B \neq 0$} is discussed in \cite{mythesis} and subject of a later paper. 

Independent of the coordinate choice $(x^{i})$ on the maximal slices, time at infinity is still measured by $\tau$. 
So the task is to find a radial coordinate transformation of the form \hbox{$r = r(C(\tau),x)$} to a line element with zero shift $B$, 
\bea
\label{eq:4mshift}
      ds^{2} & = & (-\alpha^{2}+\frac{\beta^{2}}{\gamma})\:d\tau^{2} 
               + 2\beta\:d\tau dr 
               + \gamma\:dr^2 
               + r^{2}\:d\Omega^{2} \nonumber \\ 
             & := & -A^{2}\:d\tau^{2} 
               + G\:dx^2 
               + r^{2}\:d\Omega^{2}.
\eea
Here, \hbox{$\frac{\partial r}{\partial \tau} = \frac{\partial r}{\partial C} \frac{dC}{d\tau}$} has been used, and the lapse $A$ and the 3-metric $G_{ij}$ with its radial component $G$ as functions of $\tau$ and $x$ have been introduced and are given by 
\bea
\label{eq:defineABG}
	A(\tau,x) & = & \alpha(\tau,r(\tau,x)) , \nonumber \\
	G(\tau,x) & = & \gamma(\tau,r(\tau,x)) \left( \frac{\partial r}{\partial x} \right)^{2}. 
\eea
To allow for a direct comparison with numerical puncture evolutions as produced by Cactus, the 3-metric is written in the form
\beq
\label{eq:3mshift}
	^{(3)}ds^{2} = G\:dx^2 + r^{2}\:d\Omega^{2} 
                     = \Psi^{4} \left[ g\:dx^2 + \frac{r^{2}}{\Psi^{4}}\:d\Omega^{2}\right] ,
\eeq
obtained by rescaling the 3-metric with the conformal factor
\beq
\label{eq:conformalpsi4}
	\Psi^{4} (x)  = \left((1+\frac{M}{2x})^{2} - \frac{Q^{2}}{4x^{2}} \right)^{2}.
\eeq
Here the coordinate $x$ of puncture evolutions is given in terms of the 3D Cartesian coordinates \hbox{$\left\{ X,Y,Z \right\}$} by \hbox{$x = \sqrt{X^2+Y^2+Z^2}$}, which measures the Euclidean distance to the puncture at the origin, and
\beq
\label{eq:gijandg}
	\displaystyle
	g_{ij}(\tau,x) := \frac{G_{ij}(\tau,x)}{\Psi^{4}(x)} 
        \ \ \begin{rm}{with}\end{rm} \ \ 
	\displaystyle
        g(\tau,x) := \frac{G(\tau,x)}{\Psi^{4}(x)}
\eeq
has been introduced.

Again it is worth pointing out that for the initial slice the radial grid coordinate $x$ is identical to the Schwarzschild isotropic coordinate $R$ defined in (\ref{eq:rofR}) and hence related to the Schwarzschild radial coordinate $r$ by 
\beq
\label{eq:rinitials}
	r(\tau = 0,x) = x \Psi^{2}(x)
\eeq
as discussed in Sec.~\ref{sec:oddlapse}. 
Note, furthermore, that for black hole puncture initial data the 3-metric $G_{ij}$ is implemented in isotropic coordinates, hence
\beq
\label{eq:ginitials}
	G(\tau = 0,x) = \Psi^{4}(x)
        \ \ \begin{rm}{and}\end{rm} \ \ 
	g(\tau = 0,x) = 1
\eeq
has to hold initially.

Notice that as a time-independent, trivial case the odd boundary conditions are included and lead to the isotropic lapse of Sec.~\ref{sec:oddlapse} together with the formulas (\ref{eq:rinitials}) and (\ref{eq:ginitials}) valid at all times, which yields the static Reissner-Nordstr\"om metric in isotropic coordinates. 
However, due to the negative values of the lapse in the region between the puncture and the throat, with an observer going backward in time, its numerical implementation has been found to be unstable in at least one example \cite{Brandt94}. 

In the following we use the observation that the puncture data line element for zero shift has to have a 3-metric with a time-independent determinant. 
As remarked in Sec.~\ref{sec:oddlapse}, this is a consequence of maximal slicing, \hbox{$K \equiv 0$}, in combination with a vanishing shift, \hbox{$B \equiv 0$}.
Considering the ansatz (\ref{eq:3mshift}), one can see that due to the time-independence of the conformal factor the determinant of the rescaled 3-metric \hbox{$g_{ij} = \frac{G_{ij}}{\Psi^{4}(x)}$} also has to be time-independent and hence to be a function of $x$ only. 
The latter can be determined on the initial slice to be given by \hbox{$\det{\left\{ g_{mn} \right\}} = x^{4} \sin^{2}{\theta}$} using the initial conditions (\ref{eq:rinitials}) and (\ref{eq:ginitials}).
So it follows that the rescaled radial part of the 3-metric can be written as
\beq
\label{eq:g}
	g (C(\tau),x) = \frac{x^{4} \Psi^{8}(x)}{r^{4}(C(\tau),x)}.    
\eeq
It is from \hbox{$G = \Psi^{4} g = \frac{x^{4}\Psi^{12}}{r^{4}} = \gamma \left( \frac{\partial r}{\partial x} \right)^{2}$} now trivial to infer for fixed slice label $C$ the ODE
\beq
\label{eq:drdx}
	\frac{\partial r}{\partial x} 
            = \pm \frac{\sqrt{p_{C}(r)}}{r^{4}} x^{2} \Psi^{6}(x),
\eeq
which relates the Schwarzschild radial coordinate $r$ to the radial coordinate $x$ for puncture evolutions. 
This can be integrated as
\beq
\label{eq:ximplicitofr}
	\int_{r_{C}}^{r} \frac{y^{4}\:dy}{\sqrt{p_{C}(y)}} 
  = \pm \int_{x_{C}}^{x} y^{2} \Psi^{6}(y) \:dy
\eeq
using the throat as lower integration limit.
Here $x_{C}$ denotes the location of the throat on the grid as a function of time and the ``$+$'' and ``$-$'' sign applies for the right-hand or left-hand side of the throat, respectively. 
One may readily check that for \hbox{$C = 0$} the ODE (\ref{eq:drdx}) coincides with the ODE (\ref{eq:drdR}) found for Schwarzschild isotropic coordinates leading in that case to \hbox{$x = R$} with the throat at \hbox{$r_{C = 0} = r_{+}$} and \hbox{$x_{C = 0} = x_{+} = R_{+} = \frac{1}{2} \sqrt{M^{2} - Q^{2}}$}. 
For later times, though, $x_{C}$ has to be found from (\ref{eq:ximplicitofr}) by demanding that the coordinate transformation \hbox{$r = r(C(\tau),x)$} is consistant with the requirement of a vanishing shift.
See \cite{mypaper2} for details and for a discussion of the late time behaviour of $x_{C}$ in the case of even and ``zgp'' boundary conditions.

\begin{figure}[!hb]
	\noindent
	\epsfxsize=85mm \epsfysize=60mm \epsfbox{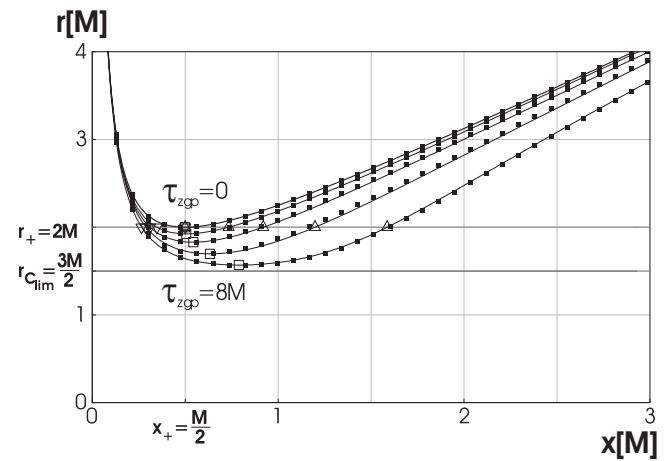} 
	\caption{
The Schwarzschild radial coordinate $r$ as given by (\ref{eq:ximplicitofr}) in the context of ``zgp'' boundary conditions is shown as a function of the isotropic radial coordinate $x$ for maximal slices with times at infinity \hbox{$\tau_{zgp} = \{0,1M,2M,4M,8M\}$}.
The slices are horizon-penetrating and approach the limiting slice \hbox{$r =$ $^{0}r_{C_{lim}} = \frac{3M}{2}$} asymptotically.
Data points from a Cactus run have been plotted as filled boxes, which agree well with analytic results.
The throat (located as minimum of the radial coordinate and depictured by a box) is found to move to the right-hand side of its initial location \hbox{$^{0}x_{+} = \frac{M}{2}$}, whereas the left-hand and right-hand event horizon (both now corresponding to \hbox{$r = $ $^{0}r_{+} = 2M$} and denoted by downward and upward pointing triangles) move to the left and right, respectively.}
\label{fig:rofx}	
\end{figure}

\begin{figure}[!ht]
	\noindent
	\epsfxsize=85mm \epsfysize=65mm \epsfbox{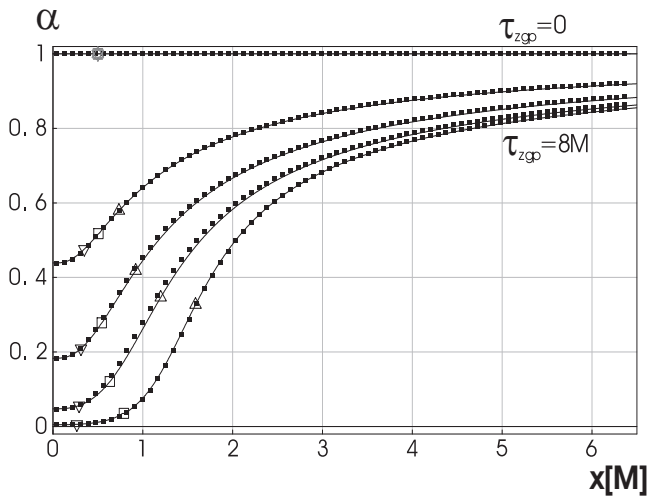} 
	\caption{
The analytically derived collapse of the lapse is shown for ``zgp'' boundary conditions as solid lines for the slice again corresponding to \hbox{$\tau_{zgp}=\{0,1M,2M,4M,8M\}$}. 
These curves have been obtained by evaluating $\alpha_{zgp}^{\pm}(C,r)$, (\ref{eq:alphazgpfinal}), in the context of the coordinate transformation (\ref{eq:ximplicitofr}) shown in Fig.~\ref{fig:rofx}. 
Cactus results from the best resolved run are shown as boxes lying almost on top of the analytically found collapsing lapse profile. 
An ``outward moving shoulder'' is obtained as the lapse decreases exponentially to zero at the puncture, at the left-hand event horizon, and at the throat, but approaches the finite value \hbox{$\frac{^{0}C_{lim}}{^{0}r_{+}^{2}} = \frac{3}{16}\sqrt{3} \approx 0.3248$} at the right-hand event horizon. 
}
\label{fig:alphax}	
\end{figure}

\begin{figure}[!ht]
	\noindent
	\epsfxsize=85mm \epsfysize=60mm \epsfbox{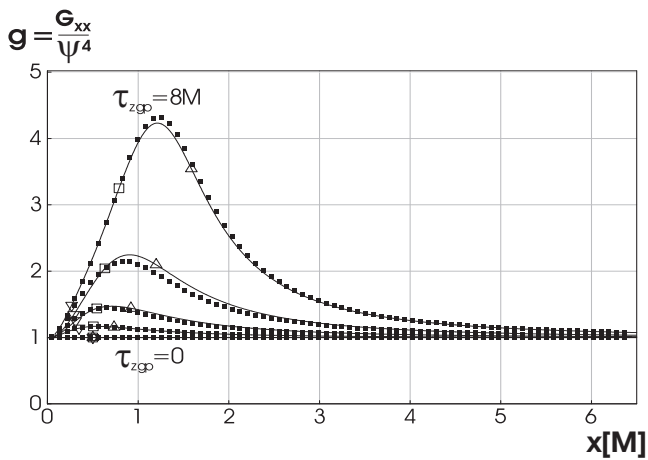} 
	\caption{
For the radial component $g$ of the conformally rescaled 3-metric $g_{ij}$ for the ``zgp'' Schwarzschild case in the manner of the previous figures, a comparison of the analytical result \hbox{$g (C,x) = \frac{x^{4} \Psi^{8}(x)}{r^{4}(C,x)}$}, (\ref{eq:g}), and Cactus output is made. 
Note that whereas \hbox{$g \equiv 1$} holds for the odd case, for other boundary conditions similar slice stretching effects are present in the radial metric component as will be discussed further in \cite{mypaper2}.
}
\label{fig:gxxx}
\end{figure}

Given these coordinates, we can finally compare the analytically given \hbox{4-metric} with Cactus output for maximal slicing of a Schwarzschild black hole with ``zgp'' boundary condition and vanishing shift. 
The time-dependent profiles, namely the coordinate transformation relating $r$ and $x$, the ``outward moving shoulder'' obtained for the collapse of the lapse and the growing peak in the radial component of the \hbox{3-metric}, are plotted in Fig.~\ref{fig:rofx}, \ref{fig:alphax}, and \ref{fig:gxxx}, respectively. 
Similar numerical results can also be found in the literature, see e.g.\ Fig.~3.16 in \cite{Camarda97} or Figs.~2 and 6 in \cite{Bernstein94} and Fig.~2 in \cite{Anninos95}, where instead of $x$ in the latter references usually a logarithmic radial coordinate has been used.

For Figs.~\ref{fig:rofx} to \ref{fig:gxxx} the reader should bear in mind that for odd boundary conditions the time-independent coordinate transformation relating Schwarzschild radial and isotropic coordinates is given by (\ref{eq:rofR}), the isotropic lapse by (\ref{eq:alphaisotropic}) and the rescaled radial part of the 3-metric is simply \hbox{$g = \frac{G}{\Psi^{4}} \equiv 1$}.
Although both the evaluation of the underlying analytical expressions (as described in \cite{mythesis}) on the one hand and the numerical simulations (with errors introduced by finite differencing and the outer boundary) on the other hand contain numerical errors, the Cactus data points virtually lay on top of the analytic curves. 

We want to emphasize that for arbitrary boundary conditions the coordinate transformation \hbox{$r = r(C,x)$}, (\ref{eq:ximplicitofr}), holds, with differences in the time-dependence of the slice label $C(\tau)$ and the (in general time-dependent) location of the throat $x_{C}$.
For this - an exception being the time-independent odd case discussed previously - e.g.\ the unbounded growth of the radial metric function \hbox{$g (C,x) = \frac{x^{4} \Psi^{8}(x)}{r^{4}(C,x)})$}, (\ref{eq:g}), as shown in Fig.~\ref{fig:gxxx} for ``zgp'' boundary conditions, is not only a feature of the puncture lapse.
Even worse, in the context of zero shift this slice stretching effect is a fundamental property of maximal slicings when avoiding singularities as will be discussed further in \cite{mypaper2}.

The development of the peak in the metric shown in Fig.~\ref{fig:gxxx} is the combined effect of local observers falling into the black hole and the collapse of the lapse function shown in Fig.~\ref{fig:alphax}. 
The coordinate points at smaller values of the isotropic coordinate have larger infall speeds, causing the radial metric component $g$ to increase towards smaller $x$.
However, at the same time, there is a competing effect due to the use of the singularity avoiding slicing. The motion of the grid points close to the puncture at \hbox{$x = 0$} is frozen as the lapse collapses there.
Well inside, the latter effect dominates, and $g$ cannot increase in time. 
This causes the radial metric function to develop a peak at a place slightly inside the right-hand event horizon. 
At this location, the difference in the infalling speed of the grid points is large, but the lapse has not completely collapsed. 

Finally, we want to mention that if one demands a time-independent 3-metric $G_{ij}$, by looking at the angular part of both $G_{ij}$ and $\gamma_{ij}$ given by $r^{2} d\Omega^{2}$ it is obvious that the coordinate transformation relating $r$ and $x$ has to be time-independent and hence to be given by (\ref{eq:rinitials}). Therefore this requirement inevitable leads to the isotropic lapse.

\pagebreak

\vspace*{0.35cm}

\vspace*{0.35cm}

\vspace*{0.35cm}

This observation is relevant for the construction of initial data for black hole punctures on a maximal slice. 
Such data can satisfy some of the necessary conditions for the existence of helical Killing vectors \cite{Tichy03a}, which is appropriate when looking for binary black holes in a quasi-equilibrium orbit.
As initial condition on the gauge one may want to impose not only the condition that the lapse is a maximal slicing lapse, but furthermore that it is everywhere positive. But as we have just argued, for Schwarzschild it is only the isotropic lapse for which the metric $G_{ij}$ is time-independent, and conversely we conclude that for positive lapse the metric will not be time-independent, which conflicts with the quasi-equilibrium condition. This was first noted in \cite{mythesis,Tichy03a} and independently in \cite{Hannam:2003tv}. 


\section{Conclusion and Outlook}
\label{sec:conclusion}

We have introduced the ``zero gradient at the puncture'' boundary condition for maximal slicing of the Schwarzschild and Reissner-Nordstr\"om spacetimes. 
Comparing analytical and numerical results for Schwarzschild, this boundary condition leads to convincing agreement with the numerically computed evolutions for maximal slicing and vanishing shift.
As an application of the analytical form for the ``zgp'' or puncture lapse, we can derive a late time limit for the lapse at the right-hand event horizon, which is consistent with previous numerical estimates.

Let us mention several directions for further investigations. First of all, in \cite{mypaper2} we will give a detailed late time analysis for the maximal slicing of Reissner-Nordstr\"om puncture data, extending the results for Schwarzschild obtained in \cite{Beig98}. 
We will discuss there the late time behaviour of the \hbox{4-metric} at the puncture, the throat and the left-hand and right-hand event horizon. 
Note that all of our numerical results refer to the Schwarzschild spacetime. 
Future numerical work could include an electric charge in the range \hbox{$0 < \mid Q \mid < M$}.

Maximal slicing is a special case of constant mean curvature slicings, see for example \cite{Gentle2001}. It would be interesting to see whether constant mean curvature slicing is amenable to a similar analysis as presented here, and whether there is a promising avenue for numerical simulations.

As mentioned in the introduction, maximal slicing has in some cases been replaced with algebraic slicings of the ``1+log'' type, see \cite{Alcubierre02a} for recent work on puncture evolutions. It may be possible to gain analytic insight into 1+log slicing, which in fact in some regards mimics maximal slicing.
For example, one would like to understand the singularity avoidance properties of 1+log slicing better. A first attempt to study this problem has met with technical difficulties, but further work is certainly warranted. 

\pagebreak

Of particular interest from the point of view of numerical relativity is an analytical study of elliptic shift conditions in Schwarzschild, say the minimal distortion shift condition.
A maximal slicing lapse together with a minimal distortion shift have been suggested as a natural coordinate choice for numerical simulations \cite{York79}. 
While the minimal distortion shift has not been successful for puncture evolutions, a closely related shift condition called Gamma freezing or conformal 3-harmonic has to some extent solved the long standing slice stretching problem of maximal slicing for single black hole and head-on collision simulations \cite{Alcubierre02a}.
In that reference this shift is implemented as an evolution equation for the shift, but it is the elliptic version which is more easily studied analytically.
A key feature of the Gamma freezing shift is that when it is combined with maximal or 1+log slicing an approximately time independent metric is obtained for the final, static black hole.
While in the present paper we answered the question whether there exists a maximal slicing corresponding to the numerically obtained ``zgp'' lapse, we now can ask the question whether for ``zgp'' maximal slices there exists a shift condition such that the evolution of the metric in the resulting coordinates is minimized. 
Some steps in this direction can already be found in \cite{mythesis}.


\bigskip
\acknowledgments
It is a pleasure to thank R.\ Beig, H.\ Beyer, \hbox{N.~\'{O} Murchadha}, E.\ Seidel, and J.\ Thornburg for helpful discussions, in particular R.\ Beig for his input during the initial stages of this work.
The numerical simulations were performed at the Albert Einstein Institute, the Max Planck Institute for Gravitational Physics.
We acknowledge the support of the Center for Gravitational Wave Physics funded by the National Science Foundation under Cooperative Agreement PHY-01-14375, in particular for a week long visit of B.\ R.\ during fall 2002. Partial support was provided by NSF grant PHY-02-18750.


\bibliographystyle{apsrev}

\bibliography{myreferences}


\end{document}